\documentclass[preprint2,twocolumn,times,tighten]{aastex63}

\usepackage{color}
\usepackage{enumitem}
\usepackage{hyperref}
\usepackage{verbatim}
\usepackage{amsmath,amstext,mathrsfs}
\usepackage[all]{hypcap} 
\usepackage{afterpage}    
\usepackage{marginnote}
\usepackage{makecell}
\usepackage{subfigure}
\usepackage{multirow}
\usepackage{lineno}

\newcommand*{\yh}{\color{red}}

\shorttitle{The Gradients Technique: Spatial Filtering Effect}
\shortauthors{Hu \& Lazarian}
\graphicspath{{./}{figures/}}

\begin{document}

\title{Tracing Magnetic Fields with the Gradient Technique: Spatial Filtering Effect and Use of Interferometers}

\email{yuehu@ias.edu; alazarian@facstaff.wisc.edu; *NASA Hubble Fellow}

\author[0000-0002-8455-0805]{Yue Hu*}
\affiliation{Institute for Advanced Study, 1 Einstein Drive, Princeton, NJ 08540, USA }
\affiliation{Department of Astronomy, University of Wisconsin-Madison, Madison, WI 53706, USA}
\author{A. Lazarian}
\affiliation{Department of Astronomy, University of Wisconsin-Madison, Madison, WI 53706, USA}

\begin{abstract}
Probing magnetic fields in astrophysical environments is both important and challenging. The Gradient Technique (GT) is a new tool for tracing magnetic fields, rooted in the properties of magnetohydrodynamic (MHD) turbulence and turbulent magnetic reconnection. In this work, we examine the performance of GT when applied to synthetic synchrotron emission and spectroscopic data obtained from sub-Alfv\'enic and trans-Alfv\'enic MHD simulations. We demonstrate the improved accuracy of GT in tracing magnetic fields in the absence of low-spatial frequencies. Additionally, we apply a low-spatial frequency filter to a diffuse neutral hydrogen region selected from the GALFA-\ion{H}{1} survey. Our results show an increased alignment between the magnetic fields inferred from GT and the Planck 353 GHz polarization measurements.
\end{abstract}


\section{Introduction}
\label{sec:intro}

Turbulent magnetic fields are key parts of understanding diffuse astrophysical environments \citep{1981MNRAS.194..809L,2004ARA&A..42..211E,2007prpl.conf...63B,MO07,CL09}. In the interstellar medium (ISM), magnetic fields are involved in regulating star formation, particle acceleration, and galaxy evolution \citep{1995ApJ...443..209A,2002RvMP...74..775W,2010ApJ...710..853C,2012ARAA...50..29,2006ApJ...647..374G,1994RPPh...57..325K,2014ApJ...783...91C,2011Natur.479..499L}. However, the magnetic field is the least-observed physical quantity involved in such processes. The primary ways of observing the magnetic fields are divided into the measurement of the plane-of-the-sky (POS) component through polarized dust emission \citep{Aetal15,2007JQSRT.106..225L} or synchrotron emission \citep{2006AJ....131.2900C,2016A&A...596A.103P,2019ApJ...871..106D}, and the line-of-sight (LOS) component from the Faraday rotation \citep{1981ApJS...45...97S,2006ApJS..167..230H,2012A&A...542A..93O} or the Zeeman splitting \citep{1976ARA&A..14....1H,2010ApJ...725..466C,2012ARAA...50..29}. The study of molecular line polarization using the Goldreich-Kulafis effect \citep{1981ApJ...243L..75G,1982ApJ...253..606G} is another way to trace magnetic fields. Line emission and absorption of atoms with fine and hyperfine structure of ground or metastable levels, i.e. Ground State Alignment (GSA, see \citealt{2006ApJ...653.1292Y,2007ApJ...657..618Y,2008ApJ...677.1401Y,2015ASSL..407...89Y,2018MNRAS.475.2415Z}) presents another promising way of exploring magnetic fields in diffuse media.

Despite having different ways to probe magnetic fields, the studies of magnetic fields remain very challenging. The measurement of dust polarization is based on the fact that aligned dust grains will produce polarized emission perpendicular to the projection of the magnetic field onto the POS \citep{2007ApJ...669L..77L,Aetal15}. There are, however, difficulties when studying the magnetic fields through dust polarimetry. For instance, the limited sensitivity of the instruments makes it difficult to measure dust polarization unless column densities of matter are sufficiently large. This presents a challenge for tracing magnetic fields at high galactic latitudes.

In this paper, we discuss a method for studying magnetic fields that is synergetic with existing methods and has several advantages. This method is the Gradient Technique (GT), which measures the gradients of observable quantities to trace the POS magnetic field. GT is developed based on the advancements in understanding MHD turbulence theory \citep{GS95} and turbulent reconnection theory \citep{LV99}. Due to fast turbulent reconnection, MHD turbulence can be presented as a superposition of turbulent eddies rotating in the direction of {\it local} magnetic field of the eddy. This property ensures the gradient of velocity fluctuations is preferential to the local magnetic field\footnote{The notion of the magnetic fields being local is not a part of the original treatment in \cite{GS95}, but this property is essential for the GT discussed in this paper. The alignment of turbulent motions with the local direction of the magnetic field has been numerically confirmed in the subsequent studies \citep{2001ApJ...554.1175M,2000ApJ...539..273C,2002ApJ...564..291C}, as well as confirmed by in situ measurements in the solar wind \citep{2016ApJ...816...15W, 2020FrASS...7...83M, 2021ApJ...915L...8D,2024ApJ...962...89Z}.} Therefore, the magnetic field orientation can be inferred from the velocity gradient rotated by 90$^\circ$.


The ability of GT to trace the magnetic field was demonstrated in the original papers (see \citealt{LY18a, PCA}). For velocity gradient, spectroscopic emission lines provide necessary velocity information \citep{LP00,LP04,KLP16,2021ApJ...910...88X,2021ApJ...915...67H,2023MNRAS.524.2994H}. More recent results of tracing magnetic field with the GT include modeling the galactic foreground polarization using neutral hydrogen data \citep{EB,2020MNRAS.496.2868L, 2020RNAAS...4..105H}, tracing the distribution of the POS magnetic fields in molecular clouds \citep{survey,velac,2020arXiv200715344A,2022MNRAS.510.4952L,2022ApJ...934...45Z,2024MNRAS.528.3897S}, the Central Molecular Zone \citep{CMZ1,CMZ2,2024MNRAS.527.1275Y}, and nearby galaxies \citep{galaxy,2023ApJ...946....8T,2023MNRAS.519.1068L,2024arXiv240402651Z}.

Due to the symmetry of velocity and magnetic fluctuations in Alfv\'{e}nic turbulence, the gradients of magnetic fluctuations are also perpendicular to the local magnetic field. As synchrotron radiation arises from relativistic electrons spiraling along magnetic field lines \citep{2006AJ....131.2900C,2016A&A...596A.103P,2019ApJ...871..106D}, the anisotropic property is also entailed by the synchrotron emission \citep{2012ApJ...747....5L}. This gave rise to the development of the magnetic field tracing based on synchrotron intensity gradients (SIGs, \citealt{2017ApJ...842...30L}) and synchrotron polarization gradients (SPGs, \citealt{2018ApJ...865...59L}). SIGs can provide a direct measurement of the POS magnetic field without the correction of Faraday rotation \citep{cluster,2024NatCo..15.1006H}. SPGs, on the other side, open a new way to study the 3D magnetic field in the Milky Way and nearby galaxies (see \citealt{2018ApJ...865...59L} for a detailed discussion).

The exploration of the response of the GT to filtering of the data was numerically performed in \citet{2017ApJ...842...30L} for SIGs and in \citet{LY18a} for Velocity Channel Gradients (VChGs). It was confirmed that the GT can restore the magnetic field using measurements of only high spatial frequencies. In this work, we provide an extensive study of the GT performance when low-spatial frequencies are removed from the images analyzed and address the question of whether such spatial filtering can improve the accuracy of the magnetic fields traced by the GT. Our study extends to multiple types of gradients, including the Intensity Gradients (IGs, see \citealt{IGs}), Velocity Centroid Gradients (VCGs, see \citealt{2017ApJ...835...41G}), VChGs \citep{LY18a}, SIGs \citep{2017ApJ...842...30L,2024NatCo..15.1006H}, and SPGs \citep{2018ApJ...865...59L}\footnote{Unlike SIGs and SPGs, which utilize synchrotron radiation, IGs, VCGs, and VChGs use spectroscopic data to study the magnetic fields.}.

The importance of this work is also related to the possibility of the analysis of the data where the low frequency information is missing, 
i.e. the interferometric data \citep{1979ASSL...76...61E,2011A&A...532A..71R,2018ApJ...865...36P}. An interferometer array can only resolve angular size $\theta$ within the range of: $\rm \lambda/A_{max}<\theta<\lambda/A_{min}$, where $\lambda$ is the observation wavelength, $\rm A_{max}$ is the longest baseline, and $\rm A_{min}$ is the minimum separation between telescopes. As for the single-dish measurement, it delivers low-spatial frequencies from zero to its characteristic angular scale approximately $\rm \lambda/D$, where $\rm D<A_{min}$ is the diameter of the telescope. In some cases, single-dish observations are missing. Therefore, it is important to know to what extent this is an impediment to the studies of magnetic fields with gradients. We numerically address this question for sub-Alfv\'enic and trans-Alfv\'enic synthetic observations.

In what follows, we illustrate the theoretical foundation of the GT in terms of MHD turbulence in \S~\ref{sec:theory}. We provide details about the numerical simulations used in this work in \S~\ref{sec:data}. In \S~\ref{sec:method}, we describe the algorithm for probing the magnetic fields through GT. In \S~\ref{sec:result}, we analyze and examine the effects of removing low spatial frequencies in GT. In \S~\ref{sec:diss} and \S~\ref{sec:conc}, we present our discussion and conclusions.

\section{MHD turbulence and gradients}
\label{sec:theory}

The GT employs the properties of MHD turbulence, in particular, the anisotropy of turbulent motions with respect to the magnetic field, i.e., the alignment of turbulent eddies with the local direction of the magnetic field. The subject has been studied over several decades \citep{1981PhFl...24..825M,1983PhRvL..51.1484M,1983JPlPh..29..525S,1984ApJ...285..109H} with the important breakthrough given by \citet{GS95}, denoted as GS95 later. There, the GS95 anisotropy scaling was suggested to characterize the anisotropy:
\begin{equation}
    k_\parallel\propto(k_\bot)^{2/3},
    \label{eq:gs}
\end{equation}
where $k_\parallel$ and $k_\bot$ are wavenumbers perpendicular and parallel to the magnetic field, respectively. There is an important caveat, however. The GS95 scaling was derived using closure relations in the global mean magnetic field reference frame. Unfortunately, this scaling is not correct in this frame of reference. Instead, in the mean magnetic field system of reference, the anisotropy is dominated by the largest eddies and is scale-independent (see \citealt{2002ApJ...564..291C}) as opposed to the prediction given by Eq.~\ref{eq:gs}. This was demonstrated, for instance, in \citet{2000ApJ...539..273C} as well as in \citet{2001ApJ...554.1175M}. 

The scale-dependent anisotropy similar to GS95 can be obtained, however, if one uses the {\it local} system of reference associated with the eddies. The eddy description of turbulence is possible due to fast 3D turbulent magnetic reconnection introduced in \citet{LV99} (henceforth LV99). Indeed, LV99 demonstrated that the turbulent motions of strongly magnetized fluid can be still eddy-like if the eddies axes are aligned with {\it local} magnetic field direction in the eddy direct vicinity. This type of motion was not considered earlier, but LV99 showed that the reconnection process reconnects the magnetic field through such an eddy within one eddy turnover. Therefore, the magnetic field cannot constrain this type of eddy motion which, in fact, provides the path of least resistance for the turbulent cascade. 

Using the eddy description of Alfv\'{e}nic turbulence it is easy to obtain its scalings. The motions perpendicular to the local magnetic field is Kolmogorov due to the absence of the back-reaction of the magnetic field. The time scales for the eddy turnover of the eddy $l_\bot/v_l$ achieves a critical balance with the period of the Alfv\'{e}n wave $l_\|/v_A$ induced by the eddies' motion. Here $l_\bot$ and $l_\|$ are the perpendicular and parallel size of eddies with respect to the {\it local} magnetic field. From the Kolmogorov scaling of velocities $v_l\sim l_\bot^{1/3}$ and the critical balance, LV99 derived the anisotropy scaling in the local reference frame:
\begin{equation}
\label{eq.lv99}
 l_\parallel\simeq L_{\rm inj}(\frac{l_\bot}{L_{\rm inj}})^{\frac{2}{3}} M_{\rm A}^{-4/3},
\end{equation}
where $M_{\rm A}$ is the ratio of the injection velocity $v_{\rm inj}$ to the Alfv\'{e}n speed $v_A$ and $L_{\rm inj}$ is the injection scale of turbulence. The universal scale-dependent anisotropy of Alfv\'{e}nic turbulence in the local magnetic field reference frame has been demonstrated in numerical simulations \citep{2000ApJ...539..273C,2002ApJ...564..291C,2001ApJ...554.1175M} and in situ measurements in the solar wind \citep{2016ApJ...816...15W, 2020FrASS...7...83M, 2021ApJ...915L...8D,2024ApJ...962...89Z}.

The picture above implies that the eddies predominately rotate in the direction aligned with the local magnetic field. It means that we can determine the magnetic field direction by detecting the eddy's preferential rotation axis, which is perpendicular to the velocity gradient. With the aforementioned Kolmogorov scaling of perpendicular motions of the eddies, the gradients of velocities scale as $v_l/l_\bot\sim l_\bot^{-2/3}$, which make the smallest resolved eddies the most prominent in the gradient signal. Those, as we discussed earlier, reveal the local magnetic field direction at the smallest scales.

The considerations above constitute the theoretical foundation for GT as applied to spectroscopic data revealing velocities of the turbulent media. Because of the symmetry of magnetic and velocity fluctuation in Alfv\'{e}nic turbulence, the gradients of magnetic fields are also perpendicular to the local direction of the magnetic field. The latter is the basis for using synchrotron emission within the GT framework \citep{2017ApJ...842...30L,2018ApJ...865...59L,2024NatCo..15.1006H}.

The situation with density fluctuations in turbulent media is somewhat more complicated (see \citealt{2007ApJ...658..423K}). The density fluctuations in subsonic turbulence are passively mixed by velocity fluctuation. Therefore, in these settings, the density gradient is also perpendicular to the magnetic fields. However, the physics becomes different in super-sonic turbulence, where shocks tend to change the density gradient's direction by 90$^\circ$ (see more discussion in \citealt{IGs,H2}). Still, as discussed in \citet{2005ApJ...624L..93B} the small amplitude density fluctuations in super-sonic turbulence still follow the GS95 scaling and their corresponding gradient can still trace the magnetic field.

Explicitly, the velocity gradient and density gradient scale as \citep{2018arXiv180200024Y}:
\begin{equation}
\label{eq.grad}
\begin{aligned}       
        \nabla \rho_l&\propto\frac{\rho_{l}}{l_\bot}\simeq\frac{\rho_0}{c}\mathscr{F}^{-1}(|\hat{k}\cdot\hat{\zeta}|)\nabla v_l,\\
        \nabla v_l&\propto\frac{v_{l}}{l_\bot}\simeq \frac{v_{\rm inj}}{L_{\rm inj}}(\frac{l_{\perp}}{L_{\rm inj}})^{-\frac{2}{3}} M_{\rm A}^{\frac{1}{3}},
\end{aligned}
\end{equation}
because the anisotropic relation indicates $l_\bot \ll l_\parallel$. Here $\rho_0$ is the mean density, $\hat{\zeta}$ is the unit vector for the Alfv\'{e}nic mode, fast mode, or slow mode, and $c$ is the propagation speed of corresponding mode. In this case, the eddies on a small scale provide the most important contribution for the gradients.

\begin{table}
\centering
\label{tab:sim}
\begin{tabular}{| c | c | c | c | c |}
\hline
Model & $M_{\rm S}$ & $M_{\rm A}$ & Resolution & $\beta$\\ \hline \hline
A1 & 0.66 & 0.12 & $792^3$ & 0.07\\
A2 & 0.63 & 0.34 & $792^3$ & 0.58\\
A3 & 0.62 & 0.56 & $792^3$/$512^3$/$480^3$ & 1.63\\
A4 & 0.60 & 0.78 & $792^3$ & 3.38\\
A5 & 0.60 & 1.02 & $792^3$ & 5.78\\  
A6 & 1.27 & 0.50 & $792^3$ & 0.31\\
A7 & 5.96 & 0.31 & $792^3$ & 0.005\\
A8 & 10.81 & 0.23 & $792^3$ & 0.001\\
A9 & 11.11 & 0.37 & $792^3$ & 0.002\\  
A10 & 10.53 & 0.46 & $792^3$ & 0.004\\
A11 & 10.61 & 0.64 & $792^3$ & 0.007\\
\hline
\end{tabular}
\caption{Description of our MHD simulations. $M_{\rm S}$ and $M_{\rm A}$ are the instantaneous values at each the snapshots are taken. The compressibility of turbulence is characterized by $\beta=2(\frac{ M_{\rm A}}{ M_{\rm S}})^2$.}
\end{table}

\section{Numerical Data}
\label{sec:data}

We perform 3D MHD simulations through ZEUS-MP/3D code \citep{2006ApJS..165..188H}, which solves the ideal MHD equations in a periodic box:
\begin{equation}
    \begin{aligned}
      &\partial\rho/\partial t +\nabla\cdot(\rho\Vec{v})=0,\\
      &\partial(\rho\Vec{v})/\partial t+\nabla\cdot[\rho\Vec{v}\Vec{v}+(p+\frac{B^2}{8\pi})\Vec{I}-\frac{\Vec{B}\Vec{B}}{4\pi}]=f,\\
      &\partial\Vec{B}/\partial t-\nabla\times(\Vec{v}\times\Vec{B})=0,
    \end{aligned}
\end{equation}
where $f$ is a random large-scale driving force, $\rho$ is the density, $\Vec{v}$ is the velocity, and $\Vec{B}$ is the magnetic field. We also consider a zero-divergence condition $\nabla \cdot\Vec{B}=0$, and an isothermal equation of state $p = c_s^2\rho$, where $p$ is the gas pressure. We use single fluid, operator-split, solenoidal turbulence injections at spatial scale $k$ = 2, and staggered grid MHD Eulerian assumption. 

\begin{figure}[t]
    \centering
    \includegraphics[width=1.0\linewidth,height=0.82\linewidth]{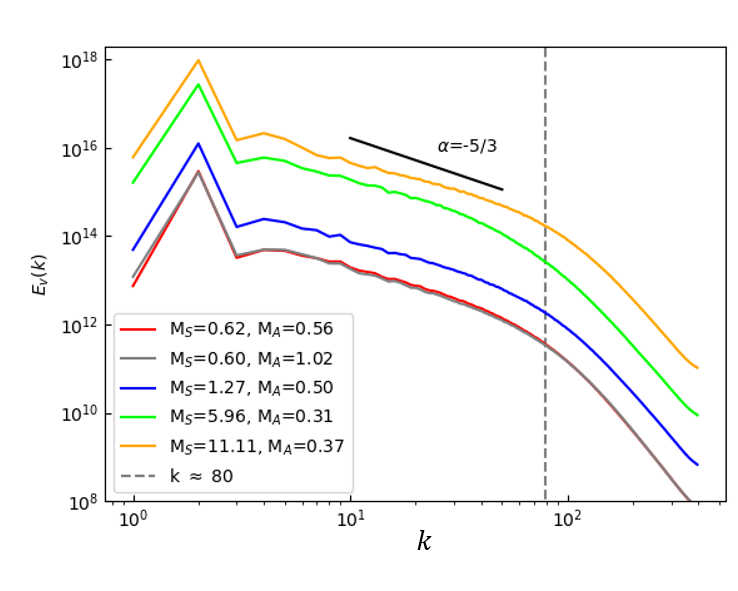}
    \caption{\label{fig:Espectrum} Energy spectra of velocity. The vertical dashed lines are used to mark locations of the dissipation scale.}
\end{figure}

To emulate a part of an interstellar cloud, we use the barotropic equation of state, i.e., these clouds are isothermal with temperature T = 10.0 K, sound speed $c_s$ = 187 m/s and cloud size $L$ = 10 pc. The sound crossing time $t_v = L/c_s$ is $\sim 52.0$ Myr, which is fixed owing to the isothermal equation of state. We vary the sonic Mach number $M_{\rm S}=v_{\rm inj}/c_{s}$ and Alfv\'{e}nic Mach number $M _{\rm A}=v_{\rm inj}/v_{A}$ to explore different physical conditions. Here $v_{\rm inj}$ is the injection velocity and $v_{A}$ is the Alfv\'{e}nic velocity. 
When $ M_{\rm S}>1$, the velocity of fluctuation is larger than the sound speed so that shocks begin appearing. We refer to the simulations in Tab.~\ref{tab:sim} by their model name or parameters. For instance, in Fig.~\ref{fig:Espectrum} we plot the energy spectral of velocity using simulations A3, A5, A6, A7, and A8. The energy spectral within the inertial range follows the Kolmogorov scaling $E_v(k) \propto k^{-5/3}$ for sub-sonic, trans-sonic, and super-sonic conditions. The dissipation scale is approximately at $k$ = 80 in our simulations. Similar simulations have been used in \citet{SFA}.

\section{Methodology}
\label{sec:method}
\subsection{The Gradient Technique}
The GT consists of multiple types of gradients: intensity gradients (IGs, see \citealt{IGs}), velocity centroid gradients (VCGs, see \citealt{2017ApJ...835...41G}), velocity channel gradients (VChGs, see \citealt{LY18a}),  synchrotron intensity gradients (SIGs, see \citealt{2017ApJ...842...30L}), and synchrotron polarization gradients (SPGs, see \citealt{2018ApJ...865...59L}), as well as synchrotron derivative polarization gradients (SDPGs, see \citealt{2018ApJ...865...59L}). 

IGs, VCGs, and VChGs are usually calculated from the spectroscopic data, for instance, the atomic \ion{H}{1} 21 cm emission line, the molecular CO emission lines, etc. This implementation is based on the theory of intensity fluctuations in Position-Position-Velocity (PPV) space pioneered in \citet{LP00} and later in \citet{KLP16}. The theory explored the possibility of using the statistics of intensity fluctuations in PPV cubes to study turbulence. The comparison of the moment-0, moment-1, and moment-2 maps calculated from PPV cubes was performed in \citet{H2}. The moment-0 map (also called intensity map, $I(x,y)$) is used to calculate IGs and the moment-1 map (also called velocity centroid map, $C(x,y)$) is used to calculate VCGs:
\begin{equation}
\label{eq:IC}
    \begin{aligned}
         I(x,y)&=\int \rho(x,y,v)dv,\\
         C(x,y)&=\frac{\int v\rho(x,y,v)dv}{\int \rho(x,y,v)dv},\\
    \end{aligned}
\end{equation}
where $\rho$ is the gas volume density, $v$ is the radial velocity along LOS. The earlier numerical and analytical studies of centroids are performed in \citet{2005ApJ...631..320E} and \citet{KLP16}, respectively. 

As for the VChGs, the calculation involves the velocity thin channel map $Ch(x,y)$:
\begin{equation}
\label{eq:Ch}
    \begin{aligned}
        Ch(x,y)&=\int_{v_0-\Delta v/2}^{v_0+\Delta v/2}\rho(x,y,v) dv,
    \end{aligned}
\end{equation}
where $v_0$ is the velocity corresponding to the maximum intensity in PPV's spectral and $\Delta v$ is the channel width $\Delta v$ satisfies:
\begin{equation}
\label{eq1}
\begin{aligned}
    \Delta v<\sqrt{\delta (v^2)},\\
\end{aligned}
\end{equation} 
here $\sqrt{\delta (v^2)}$ is the velocity dispersion on the scales that turbulence is studied. Eq.~\ref{eq1} was proposed in \citet{LP00} as a crucial criterion to distinguish the dominance of velocity fluctuation in PPV cubes. Once the channel width satisfies with Eq.~\ref{eq1}, the intensity fluctuations in PPV cubes could arise due to turbulent velocities along the LOS. This is called the velocity caustics effect, which must inevitably present in the thin velocity channel maps, as a natural result of non-linear mapping from the real space to the PPV space. In what follows, we use the description in \citet{LP00} that is provided for the central channels. In a real scenario, the PPV cubes of emission lines are measured in terms of radiation temperature $T_R(x,y,v)$, or brightness temperature, rather than gas volume density. The gas volume density used in Eq.~\ref{eq:IC} and Eq.~\ref{eq:Ch} should be replaced by $T_R(x,y,v)$ accordingly. In the case of optically thick emission, Eq.~\ref{eq:IC} and Eq.~\ref{eq:Ch} only sample the outskirts of the cloud.
 
As for SIGs and SPGs, the calculation utilizes the intensity of synchrotron emission $S(x,y)$ and the polarized synchrotron intensity $P(x,y)$ given by Stokes parameters $Q(x,y,z)$ and $U(x,y,z)$ :
\begin{equation}
\label{eq.sp}
    \begin{aligned}
      S(x,y)&\propto\int  n_e B_\bot^\gamma(x,y,z) dz\\
      P(x,y)&= |\int (Q+iU) dz|,\\
      Q(x,y,z)&\propto p n_e (B_y^2-B_x^2),\\
      U(x,y,z)&\propto p n_e (-2B_xB_y),
    \end{aligned}
\end{equation}
where $B_\bot = \sqrt{B_x^2 + B_y^2}$ corresponds to the magnetic field component perpendicular to the LOS, i.e., the z-axis. $p$ is the polarization fraction, which is to be assumed constant, and $n_e$ is the relativistic electron density. We used $\gamma=2$ instead of the actual fractional power of the index appealing to the quantitative theory in \citep{2012ApJ...747....5L}. There the relations between the power spectra and structure functions of synchrotron emission for arbitrary $\gamma$ were related to the statistics obtained for $\gamma=2$ through the function of $\gamma$ defined there. Therefore results obtained for $\gamma=2$ can trivially be generalized for an arbitrary $\gamma$. Similar to the case of molecular emission, $S(x,y)$ measures the full magnetic field structure when the emission is optically thin.
\begin{figure*}[t]
    \centering
    \includegraphics[width=0.9\linewidth,height=0.9\linewidth]{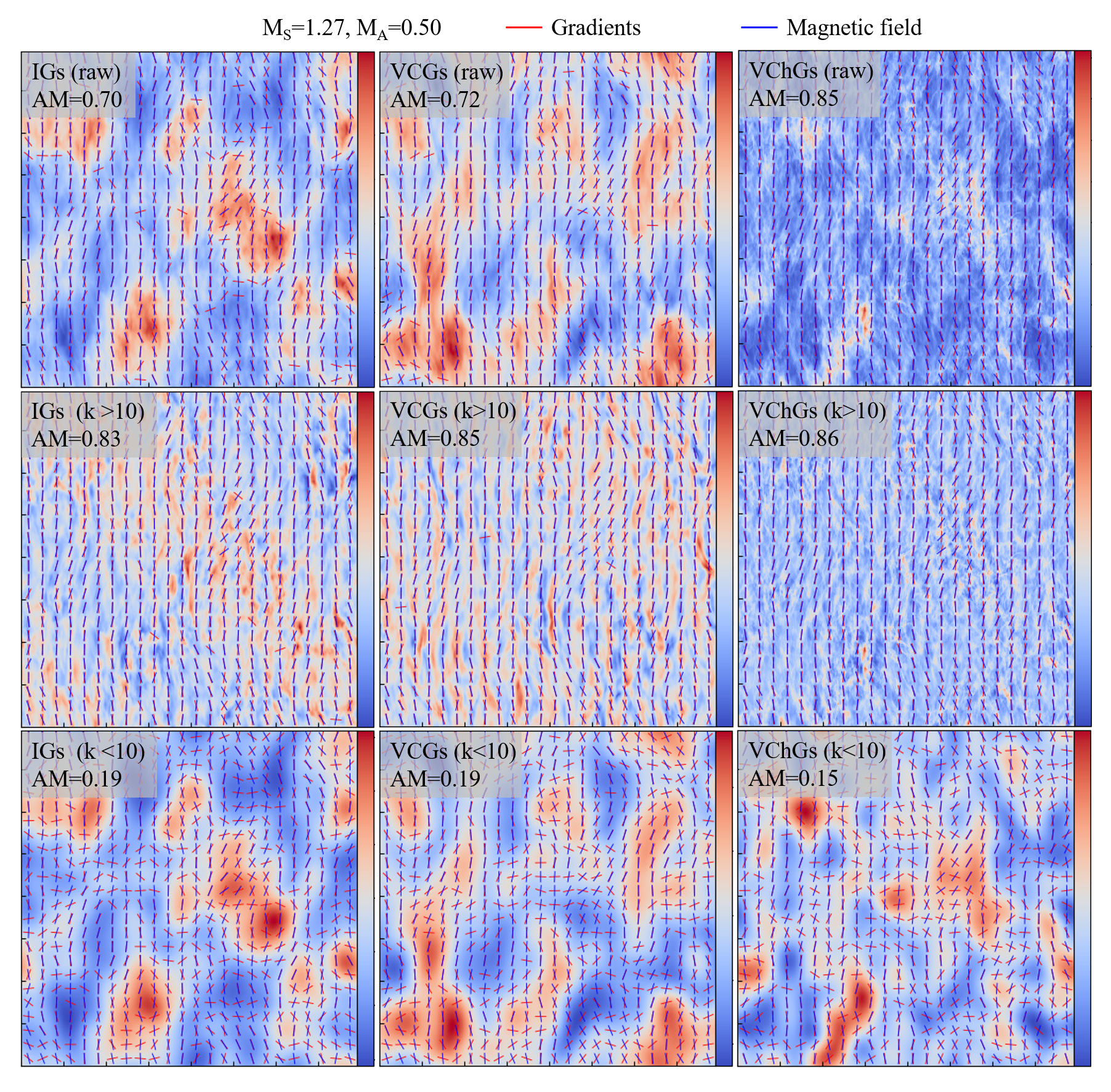}
    \caption{\label{fig:map} The plot of the magnetic field orientation probed by GT (red) and synthetic dust polarization (blue) with sub-block size 33 pixels. The plot is overlaid on the intensity maps (the 1st column), velocity centroid maps (the 2nd column), and velocity channel maps (the 3rd column). The gradients are calculated within full spatial frequency range (the 1st row), $k>10$ (the 2nd row), and $k<10$ (the 3rd row), respectively.}
\end{figure*}

For the sake of simplicity, while addressing the SPGs, we discuss only a very special case of no Faraday rotation influencing the arriving polarized signal. This corresponds, for instance, to high, e.g. tens of GHz, frequencies of radio emission. The value of the SPGs, as they are introduced in \citet{2018ApJ...865...59L} is that the SPGs can study magnetic fields at any frequency and by combining different frequencies one can restore the 3D structure of the magnetic field. We will consider the effects of spatial filtering on SPGs in a general case within a separate study. 

The pixelized gradient map $\psi_{g}(x,y)$ of each 2D map is calculated from the convolution with 3 $\times$ 3 Sobel kernels $G_x$ and $G_y$\footnote{
The Sobel kernels are defined as:
$$
G_x=\begin{pmatrix} 
-1 & 0 & +1 \\
-2 & 0 & +2 \\
-1 & 0 & +1
\end{pmatrix},
G_y=\begin{pmatrix} 
-1 & -2 & -1 \\
0 & 0 & 0 \\
+1 & +2 & +1
\end{pmatrix}
$$
}:
\begin{equation}
\label{eq:conv}
\begin{aligned}
  \bigtriangledown_x f(x,y)=G_x * f(x,y),  \\  
  \bigtriangledown_y f(x,y)=G_y * f(x,y),  \\
  \psi_{g}(x,y)=\tan^{-1}(\frac{\bigtriangledown_y f(x,y)}{\bigtriangledown_x f(x,y)}),
\end{aligned}
\end{equation}
where $f(x,y)$ can be replaced by I(x,y), C(x,y), Ch(x,y), $S(x,y)$, or $P(x,y)$. $\bigtriangledown_x f(x,y)$ and $\bigtriangledown_y f(x,y)$ are the x and y components of gradient respectively. $*$ denotes the convolution. 
\begin{figure*}[t]
    \centering
    \includegraphics[width=1.0\linewidth]{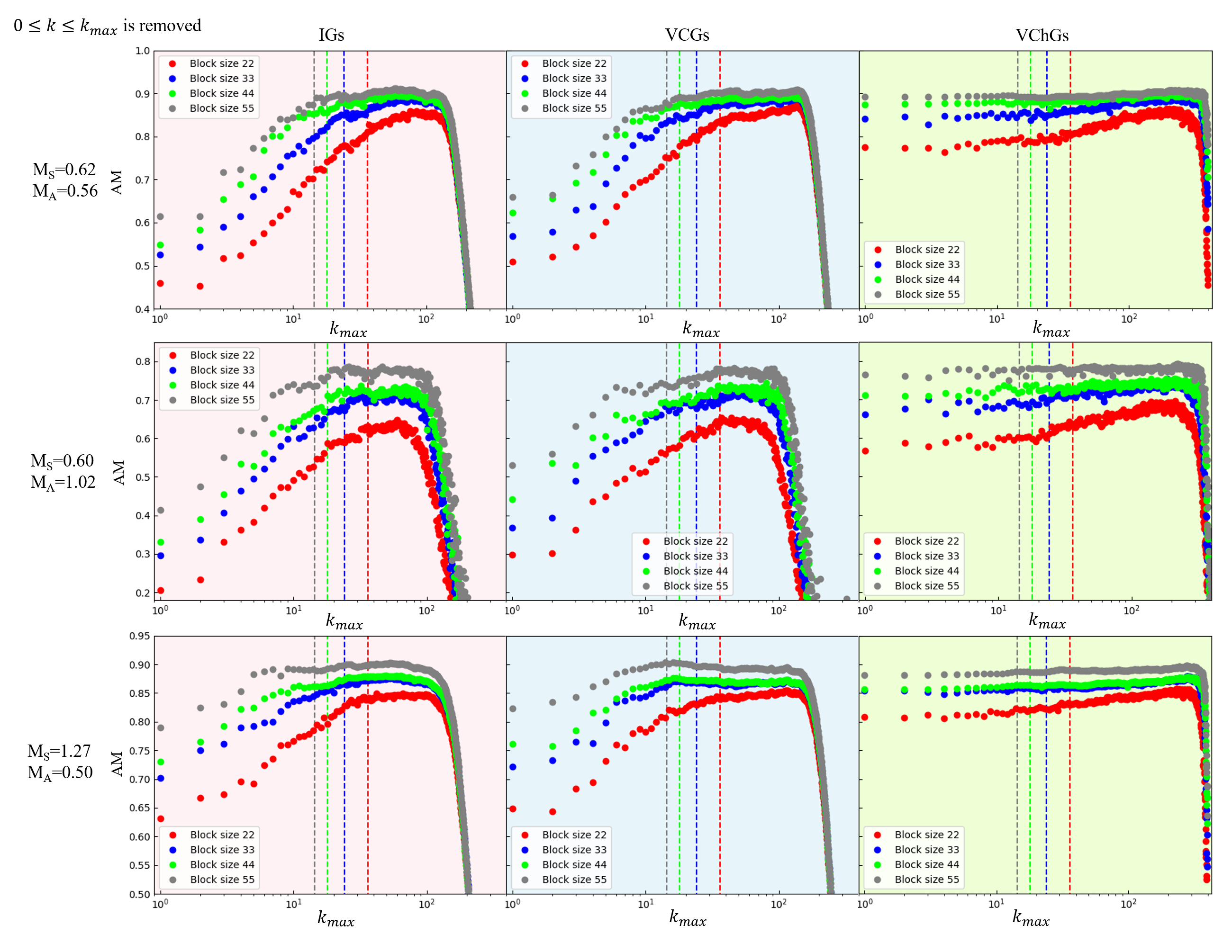}
    \caption{\label{fig:sub_kmax} The correlation of AM and $k_{max}$. $k_{max}$ (the $x$-axis) indicates the spatial frequency from $k=0$ to $k=k_{max}$ is removed. The AM is calculated from IGs (the 1st column)/VCGs (the 2nd column)/VChGs (the 3rd column) and the magnetic field inferred from polarization. The dashed lines represent corresponding the $k$ values of the selected sub-block sizes.}
\end{figure*}

\begin{figure*}[t]
    \centering
    \includegraphics[width=1.0\linewidth,height=1.0\linewidth]{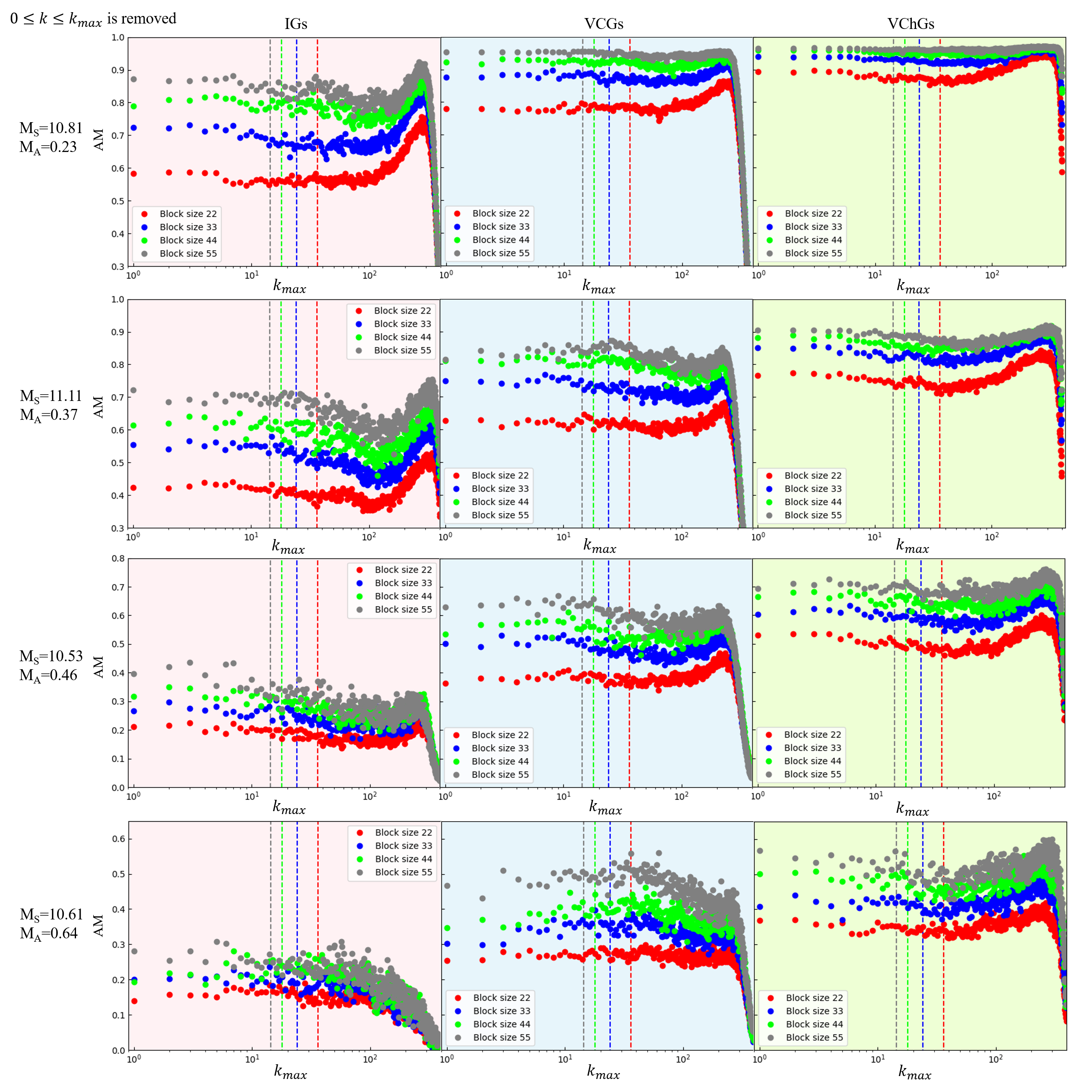}
    \caption{\label{fig:sup_kmax} The correlation of AM and $k_{max}$. $k_{max}$ (the $x$-axis) indicates the spatial frequency from $k=0$ to $k=k_{max}$ is removed. The AM is calculated from IGs (the 1st column)/VCGs (the 2nd column)/VChGs (the 3rd column) and the magnetic field inferred from polarization. The dashed lines represent corresponding the $k$ values of the selected sub-block sizes.}
\end{figure*}
\subsubsection{Sub-block averaging}
Eq.~\ref{eq:conv} produces the raw gradient map $\psi_{g}(x,y)$. However, as the anisotropy of turbulent eddies concerning the local magnetic field is a statistical concept, the individual raw gradient is not necessarily required to have any relation to the local magnetic field direction. The perpendicular relative orientation of gradients and magnetic field only appears when the gradient sampling is enough. To extract the anisotropy, the critical step after getting $\psi_{g}(x,y)$ is taking the average of gradients' orientation within a sub-block of interest, called the sub-block averaging method \citep{YL17a}.

The sub-block averaging method is implemented as: (i) draw the histogram of gradients' orientation within a sub-region; (ii) fit the histogram with the Gaussian curve and take the expectation value of the Gaussian fitting. The resulting expectation value $\psi(x,y)$ of the Gaussian distribution reflects the statistical most probable orientation of gradients. By rotating $\psi(x,y)$ with 90$^\circ$, we output the predicted local mean magnetic field. Without specification, we automatically rotate the gradient by 90$^\circ$ to align with the magnetic field in the following.

\subsubsection{K-space filter}
Several approaches to improving GT accuracy have been proposed. For instance, \citet{PCA} utilized Principal Component Analysis (PCA) to extract crucial velocity components in PPV cubes. \citet{LY18a} employed the Moving Window method, while \citet{cluster} introduced the Rotational Image test to correct abnormal gradients. In this study, we investigate the contribution of different spatial frequencies to GT and explore further possibilities for enhancing its accuracy.

The effect of removing low spatial frequencies was examined in \citet{2017ApJ...842...30L}, where it was found that super-Alfv\'enic turbulence gradients can still trace the magnetic field even when low spatial frequencies are removed. Additional studies on GT's capability to recover sub-Alfvé\'enic and trans-Alfv\'enic magnetic field maps while removing low spatial frequencies are presented in \citet{LY18a}.

Spatial filtering is performed on 2D data sets, with the capital letter $K$ denoting the 2D wavenumber. The K-space filter is implemented as follows: (i) apply the Fast Fourier Transform (FFT) to a 2D map; (ii) remove the intensity values corresponding to a wavenumber $k$ (or a range of $k$) of interest; (iii) apply the inverse FFT to the processed map, and then perform the gradients' calculation.

\subsubsection{Alignment measure}
To quantify the relative alignment between the magnetic fields and rotated gradient angle, \citet{2017ApJ...835...41G} introduced the \textbf{Alignment Measure} (AM): 
\begin{align}
\label{AM_measure}
\rm AM=2(\langle \cos^{2} \theta_{r}\rangle-\frac{1}{2})
\end{align}
where $\theta_r$ is the angular difference of two vectors, while $\langle ...\rangle$ denotes the average within a region of interests. In the case of a perfect global alignment between rotated gradients and the POS magnetic field gradient, we get AM = 1, while AM = -1 indicates global gradients are perpendicular to the POS magnetic field. In this study, we define AM = 0.70 as a threshold of rough alignment ($\theta_r\approx22.5^\circ$) and  AM = 0.90 as a good alignment ($\theta_r\approx12.5^\circ$). The standard error of the mean gives the uncertainty $\sigma_{AM}$ which is negligible for sufficient sampling.

\section{Result}
\label{sec:result}
\subsection{Analysis of IGs, VCGs, and VChGs}
In this section, we analyze the effect of spatial frequency on the accuracy of IGs, VCGs, and VChGs. First of all, we test the case of removing low-spatial-frequency (from $k=0$ to $k_{max}$) and removing high-spatial-frequency (from $k_{min}$ to $k=792$). 

Fig.~\ref{fig:map} is an example illustrating the role of spatial frequency in 2D images. We make a comparison of the raw image, the high-spatial-frequency image ($k<10$ is removed), and the low-spatial-frequency image ($k>10$ is removed). We can see the high-spatial-frequency components correspond to small-scale structures, and low-spatial-frequency components correspond to large-scale structures. We also calculate the alignment between the gradients and the magnetic fields weighted by density. For the raw images (full spatial frequency range), IGs, VCGs, and VChGs give AM = 0.70, 0.72, and 0.85, respectively. After the low-spatial-frequency components were removed, the AM of IGs and VCGs increased to 0.83 and 0.85. When the high-spatial-frequency components are removed, however, the AM decreased to 0.19 (IGs), 0.19 (VCGs), and 0.15 (VChGs). It means the high-spatial-frequency components, i.e., small-scale structures, dominate the alignment of gradients and the magnetic field.

\subsubsection{Removing low-spatial frequency}
In Fig.~\ref{fig:sub_kmax}, we plot the AM value as a function of $k_{max}$, which means the spatial frequencies from $k=0$ to $k=k_{max}$ are removed. We vary the sub-block size from 22 pixels to 55 pixels. In the case of sub-sonic and trans-sonic turbulence, the removal of low-spatial frequencies increases the AM values. In particular, for IGs and VCGs of sub-Alfv\'{e}nic simulations ($M_{\rm A}=0.56$ and $M_{\rm A}=0.50$), the AM gets boosted to approximately 0.9 in the range of $10<k_{max}<100$ around. Also, the AM of IGs and VCGs get saturated when $k_{max}$ is larger than the corresponding $k$ values of the selected sub-block sizes. Still, the AM drops when $k_{max}$ is larger than 100, which is approximately the 3D dissipation scale of turbulence. This means the optimal sub-block should be larger than the minimum scale of turbulent fluctuations. In terms of the sup-Alfv\'{e}nic and sub-sonic case ($M_{\rm S}=0.60$, $M_{\rm A}=1.02$), the initial AM values of IGs and VCGs are 0.2 and 0.3, respectively. After removing the low-spatial frequencies, the AMs are improved to 0.7. 

As for sub-sonic/trans-sonic VChGs, the situation is slightly different. As shown in Fig.~\ref{fig:map}, the structures in the raw thin velocity channel maps are much more filamentary than the intensity map and the velocity centroid ma, due to the velocity caustic effect \citep{LP00}. These filamentary structures are dominated by the velocity fluctuations and are better aligned with the magnetic fields.\footnote{The effect of thermal broadening is not considered here as we assume that we can use heavy tracers, e.g., heavy molecules in the flow. Otherwise, one should employ the procedures of the thermal broadening removal described in \citet{2018arXiv180200024Y}.} Therefore, raw VChGs (full spatial frequency range) always give a higher AM value than raw IGs and raw VCGs. The removal of low-spatial-frequencies still slightly improves the alignment (see Fig.~\ref{fig:sub_kmax}). However, being different from IGs and VCGs, the curve of VChGs is more flat. At the same time, in all cases, the AM drops off considerably after the AM gets its maximum value. The drop-off scale is around $k_{max}=200$, larger than the dissipation scale in 3D. This can result from non-linear spectroscopic mapping from real position space to PPV space, which creates smaller-scale structures. In 2D maps, the correlation fading of turbulence at small separations is compensated by the correlations along the LOS when the LOS of two pixels is closer than the dissipation scale. Therefore, we can see the alignment at a 2D scale smaller than the dissipation scale.
\begin{figure*}[t]
    \centering
    \includegraphics[width=1.0\linewidth]{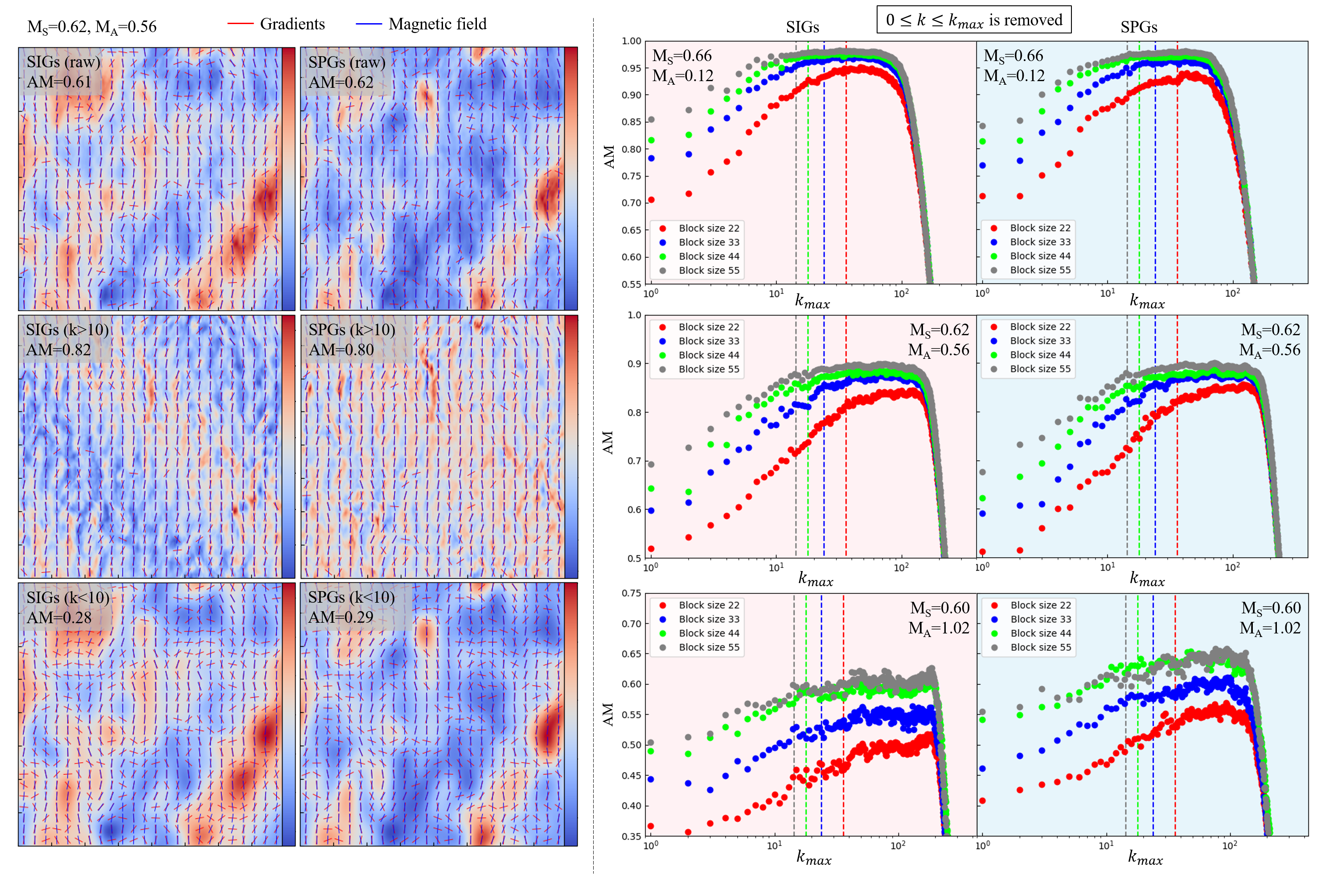}
    \caption{\label{fig:skmax} \textbf{Left:} The plot of the magnetic field orientation probed by GT (red) and synthetic polarization (blue) with sub-block size 33 pixels. The plot is overlaid on the synchrotron intensity maps (left), and synchrotron polarization maps (right). The gradients are calculated within the full spatial frequency range (the 1st row), $k>10$ (the 2nd row), and $k<10$ (the 3rd row), respectively. \textbf{Right:} The correlation of AM and $k_{max}$. $k_{max}$ (the $x$-axis) indicates the spatial frequency from $k=0$ to $k=k_{max}$ is removed. The AM is calculated from SIGs (the 1st column)/SPGs (the 2nd column) and the magnetic field inferred from polarization. The dashed lines represent corresponding the $k$ values of the selected sub-block sizes.}
\end{figure*}

As for supersonic turbulence, the presence of shocks complicates the picture of turbulent density. For instance, \citet{2005ApJ...624L..93B} revealed that the density spectrum of supersonic turbulence becomes shallower because shocks accumulate the fluid into the local and highly dense structures. In terms of gradients, their direction flips by 90$^\circ$ in front of shocks so that the AM gets decreased \citep{IGs,H2}. 

In Fig.~\ref{fig:sup_kmax}, here for the IGs of super-sonic simulations $M_{\rm S}=10.81$ and $M_{\rm S}=11.11$, we see the AM shows a depression range until $k_{max}\approx100$ which is larger the block-size scale. Different from the sub/trans-sonic cases, when $k_{max}$ is larger than the inverse block-size scale, the AM gets negative or zero improvement. We expect this come from the high-density contrast associated with shocks produced by strong fluctuations in this highly supersonic simulation. The scale of high-density contrast is smaller than the block-size scale. As a result, the high-density contrasts keep contributing negative AM even when $k_{max}$ is larger than the block-size scale. For the IGs, the depression of AM is also related to the $M_{\rm A}$. When $M_{\rm A}$ increases to 0.64, the depression is more significant and gets no improvement at larger $k_{max}$. The depression of AM also appears in VCGs and VChGs, but it is less significant than IGs. Recall that in the number of cases, e.g. in particular for low Mach numbers, the turbulent velocity contribution is dominating over the density contribution in the velocity centroid map and the thin velocity channel map (see \citet{LP00} \citealt{2005ApJ...631..320E}). This was recently demonstrated, for instance, in \citet{H2}. The shocks therefore have limited effect on VCGs and VChGs. In any case, the alignment drops off at a scale smaller than the dissipation scale. The explanation is similar to the sub-sonic case above. 
\begin{figure*}[t]
    \centering
    \includegraphics[width=1.0\linewidth]{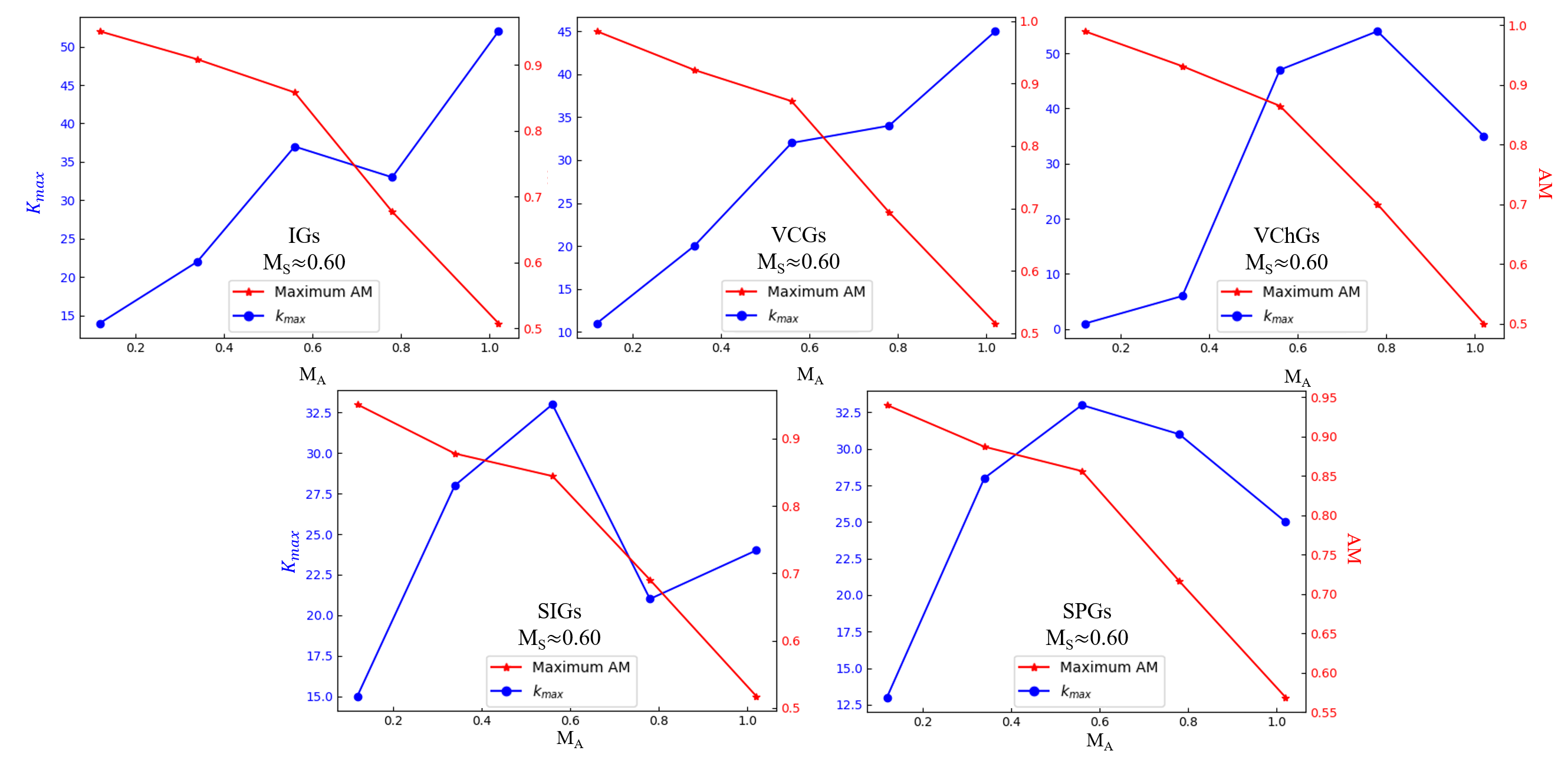}
    \caption{\label{fig:ma} \textbf{Red:} plot of the maximum AM that can be obtained from removing the spatial frequencies as a function of $M_{\rm A}$. \textbf{Blue}: the correlation of $K_{max}$ and $M_{\rm A}$. {\rm $K_{max}$ here means the maximum AM is obtained when the wavenumber from zero to $K_{max}$
    is removed.} The AM is calculated from the gradients and the magnetic field inferred from polarization using a sub-block size of 22 pixels. }
\end{figure*}

In Appendix~\ref{appendx:noise}, we study the effect of noise on the performance of GT. We introduce Gaussian noise to 2D maps I(x,y), C(x,y), and Ch(x,y) processed by the low-spatial frequency filter. The noise's amplitude in each map is 10\% of their mean intensity. We find when the sub-block size is small ($\approx 22$ pixels) the noise degrades the maximum value of AM. However, the noise's contribution can be canceled out by increasing the sub-block size. Also, we analyze the numerical effect coming from the finite resolution of the simulation. We use simulation A3 ($M_{\rm S}=0.62$, $M_{\rm A}=0.56$) with resolution 792$^3$, 512$^3$, and 480$^3$. We find in the cases of low resolution, the AM of gradients, and the magnetic field weighted by density is still improved by filtering out the low-spatial frequencies. The numerical effect has a significant effect on the dissipation scale so that the AM drops off at the smaller $k_{max}$ when the resolution is low. 

In Appendix~\ref{appendx:gravity}, we test also the effect of self-gravity which potentially can break up the turbulence's picture. Our studies showed that in the case of gravitational collapse, as the acceleration induced by the collapse is along the magnetic field direction, density, and velocity gradients are expected to change their orientation from perpendicular to magnetic fields to align with magnetic fields \citep{YL17b,survey,Hu20}. In this case, the rotated gradients become perpendicular to the magnetic field resulting in a negative AM value. The effect of self-gravity is scale-dependent, which means the gravitational collapse can happen at a small local region where the gravitational energy surpasses the kinetic and magnetic field energy. The results in Appendix~\ref{appendx:gravity} show that in the later stage of collapsing (around 0.43 free fall time), all types of gradients (IGs, VCGs, and VChGs) exhibit smaller AM at small scales. Nevertheless, \citet{Hu20} proposed the double-peak algorithm to identify the gravitation collapsing region. Once the collapsing region is determined, by re-rotating the gradients 90$^\circ$, the alignment gets recovered.  

Also, we vary the inclination angle between the mean magnetic field and the POS in simulations (see Appendix.~\ref{appendx:gamma}). The mean magnetic field is initially perpendicular to the LOS. However, when the magnetic fields are inclined with respect to the LOS, we have to consider the projection effect for the 2D magnetic field structures. We find the inclination angle constrains the maximum value of the AM (gradients and the magnetic field weighted by density). When the inclination angle increases, the mean magnetic field tends to be parallel to the LOS and the projected magnetic fields get more irregular in the POS. Therefore, the gradients get less aligned with the magnetic field and the AM decreases with the increment of the inclination angle.

\subsection{Analysis of SIGs and SPGs}

Fluctuations of synchrotron emission can also be used to study properties of the magnetic field \citep{2012ApJ...747....5L}. For instance, \citet{2017ApJ...842...30L} and \citet{2018ApJ...865...59L} proposed to trace the magnetic field orientation using SIGs or SPGs. SIGs can directly measure the POS magnetic field direction getting rid of the correction of Faraday rotation.

In Fig.~\ref{fig:skmax}, we give an example of tracing the magnetic field through the SIGs and SPGs. Here we did not consider the effect of Faraday rotation on synchrotron polarization, which will be studied elsewhere. We make a comparison of the raw image, the high-spatial-frequency image ($k<10$ is removed), and the low-spatial-frequency image ($k>10$ is removed). Similar to Fig.~\ref{fig:map}, we calculate the alignment between the SIGs/SPGs and the magnetic fields weighted by density. For the raw images, SIGs, and SPGs give AM = 0.61, 0.62, respectively. After the low-spatial frequency components are removed, the AM of SIGs and  SPGs get increased to 0.82 and 0.80, respectively. When the high-spatial frequency components are removed, however, the AM of the three cases decreased to 0.30. It means the high-spatial-frequency components, i.e., small-scale structures, dominate the alignment of gradients and the magnetic field. The SIGs, which utilize the intensity of synchrotron emission, therefore provide a reliable measurement of the POS magnetic fields. As the SIGs are not subject to Faraday rotation, it does not require multiple frequency measurements to compensate for the Faraday effect \citep{2017ApJ...842...30L}.

For low-frequency synchrotron radiation, the effect of Faraday depolarization allows to use of SPGs at multiple frequencies and this way to provide the 3D magnetic field tracing \citep{2018ApJ...865...59L}. This effect provides the wavelength-dependent LOS depth to the polarized synchrotron intensity. In this case, only the regions close to the observer contribute to the measured polarization, while unpolarized synchrotron radiation comes from more distant regions. The condition for decorrelation of Faraday rotation is given by \citep{LP16}:
\begin{equation}
\label{eq.11}
\begin{aligned}
      \lambda^2\Phi&=0.81\lambda^2\int^{L_{eff}}_0n_{e,T}B_zdz\approx1,\\
\end{aligned}
\end{equation}
where $n_{e,T}$ and $B_z$ are the thermal electron density and the LOS magnetic fields. $\lambda$ represents the wavelength of measurements. The effective depth $L_{eff}$ represents the distance to which the polarized radiation is collected effectively. The effect of removing the low-spatial frequencies in SPGs with the presence of Faraday depolarization will be studied elsewhere. 

\subsection{The dependence on $M_{\rm A}$}
When the magnetic field is weak, the motion of turbulence at the injection scale is more similar to the hydrodynamic type due to the relatively weak back-reaction of the magnetic field. The turbulent eddies are isotropic in large scales instead of anisotropic. Therefore, it is important to study the dependence of GT on $M_{\rm A}$.

We select a set of sub-sonic simulations ($M_{\rm S}\approx0.60$) with various $M_{\rm A}$. The analysis of removing low-spatial frequencies from $k=0$ to $k=k_{max}$ is repeated for all types of gradients. In Fig.~\ref{fig:ma}, we identify the maximum AM value of gradients and the magnetic field that can be obtained in a given $M_{\rm A}$. We see the maximum AM decreases with the increase of $M_{\rm A}$. Nevertheless, the AM values are always larger than 0.60 when $M_{\rm A} \approx 1$. It means globally the gradients are still aligned with the weak magnetic fields, although relatively the alignment is less than the cases of strong magnetic fields. Also, we determine the $K_{max}$ value at which the maximum AM is obtained when the wavenumber from zero to $K_{max}$ is removed as shown in Fig.~\ref{fig:ma}.  We find the saturation scale $K_{max}$ is generally increasing with the increment of $M_{\rm A}$. In general, when the $K_{max}$ is larger than 15, the gradients show the best alignment with the magnetic field. This means high spatial frequencies ($k>15$) dominate when tracing the magnetic fields with gradients. 
\begin{figure*}[t]
    \centering
    \includegraphics[width=0.9\linewidth]{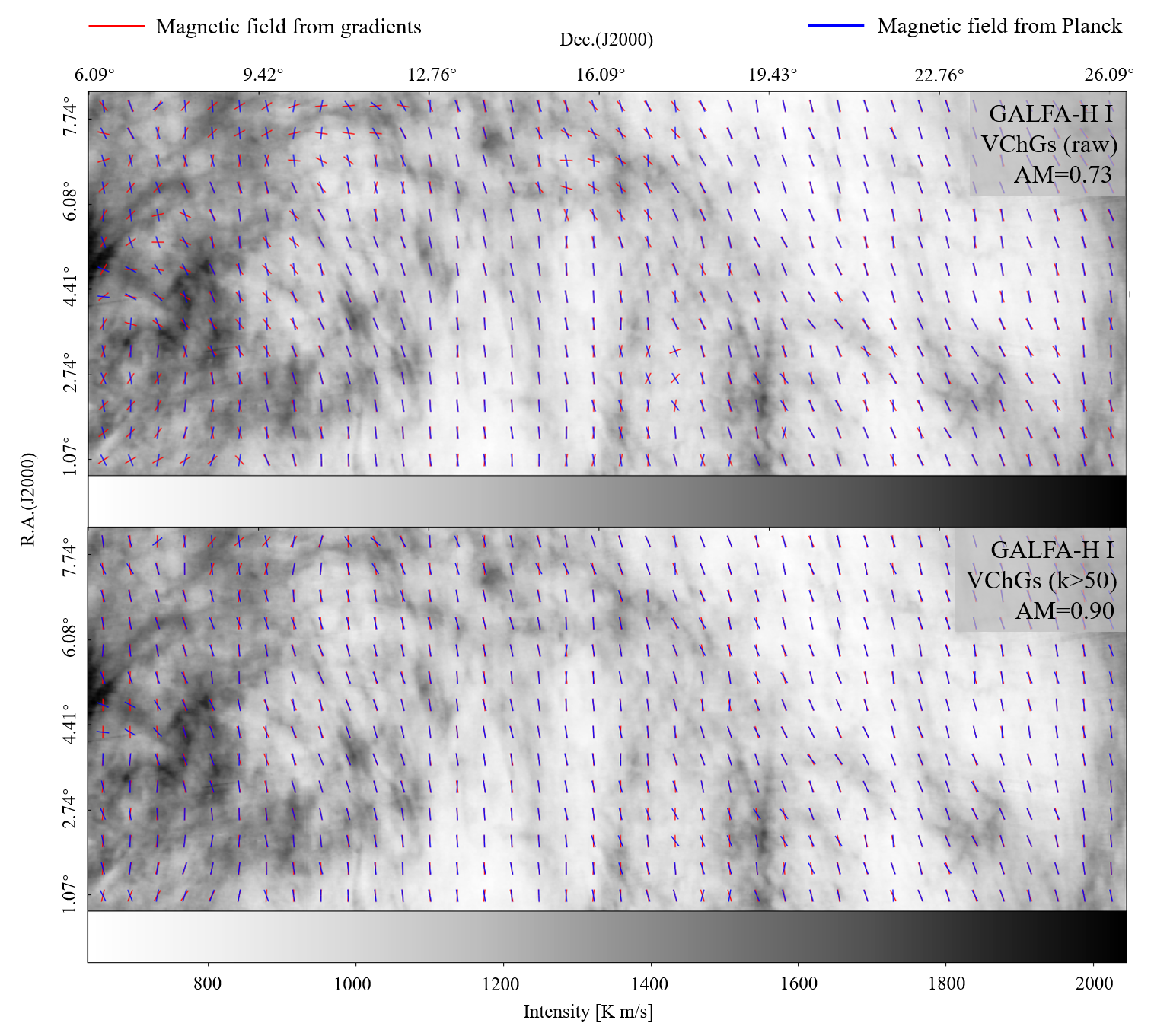}
    \caption{\label{fig:galfa} The plot of the magnetic field orientation probed by VChGs (red) and Planck polarization (blue) with resolution $\approx0.5^\circ$. The plot is overlaid on the \ion{H}{1} intensity map integrated from the velocity range of -23.0 km/s $<v<$ 14.0 km/s. \textbf{Top:} the VChGs are calculated from the GALFA-\ion{H}{1} data within full spatial frequency $k$ range. \textbf{Bottom:} the VChGs are calculated from the GALFA-\ion{H}{1} data with $k>50$.}
\end{figure*}
\subsection{Application to GALFA-\ion{H}{1} data}
To test the GT in observation, we use the high spatial and spectral resolution neutral hydrogen data from Data Release 2 (DR2) of the Galactic Arecibo L-Band Feed Array \ion{H}{1} survey (GALFA-\ion{H}{1}) with the Arecibo 305 m radio antenna \citep{2018ApJS..234....2P}. GALFA-\ion{H}{1} has a gridded angular resolution of $1'\times1'$ per pixel, a spectral resolution of 0.18 km/s, and a brightness temperature noise of 40 mK RMS per 1 km/s.

For illustration, we select a \ion{H}{1} cloud, which stretches from
R.A.$\approx-0.08^\circ$ to $8.26^\circ$ and Dec.$\approx6.09^\circ$ to $26.09^\circ$, as an example. This region spans from galactic latitude b $\approx-30^\circ$ below the Galactic Plane to b $\approx-60^\circ$, i.e., close to the Galactic south pole. We analyze the \ion{H}{1} data within velocity range -23.0 km/s $<v<$ 14.0 km/s, which contains the majority of the \ion{H}{1} emission. In Fig.~\ref{fig:galfa}, we give an example of tracing the magnetic field morphology using VChGs. We make a comparison of VChGs and the Planck 353GHz polarized dust signal data from the Planck 3rd Public Data Release (DR3) 2018 of High-Frequency Instrument \citep{2018arXiv180706207P}. The Planck observations provide Stokes parameter maps $I$, $Q$, and $U$, so the POS magnetic field orientation angle $\theta$ can be derived from the Stokes parameters: $\theta=\frac{1}{2}\tan^{-1}(U/Q)-\frac{\pi}{2}$. In this work, we smooth the $Q$, $U$ maps through a Gaussian filter with the FWHM $\approx$ 10 arcmin. The gradients are smoothed in a similar way by constructing the pseudo-Stokes parameters $\rm Q_g$ and $\rm U_g$:
\begin{equation}
    \begin{aligned}
      Q_g(x,y)=I(x,y)\cos(2\psi(x,y)),\\
      U_g(x,y)=I(x,y)\sin(2\psi(x,y)).
    \end{aligned}
\end{equation}
The Gaussian filter is applied to both $Q_g$ and $U_g$. The resulting gradient vectors are calculated from $\psi'(x,y)=\frac{1}{2}\tan^{-1}(U_g/Q_g)$.

\begin{figure}[t]
    \centering
    \includegraphics[width=1.0\linewidth,height=0.8\linewidth]{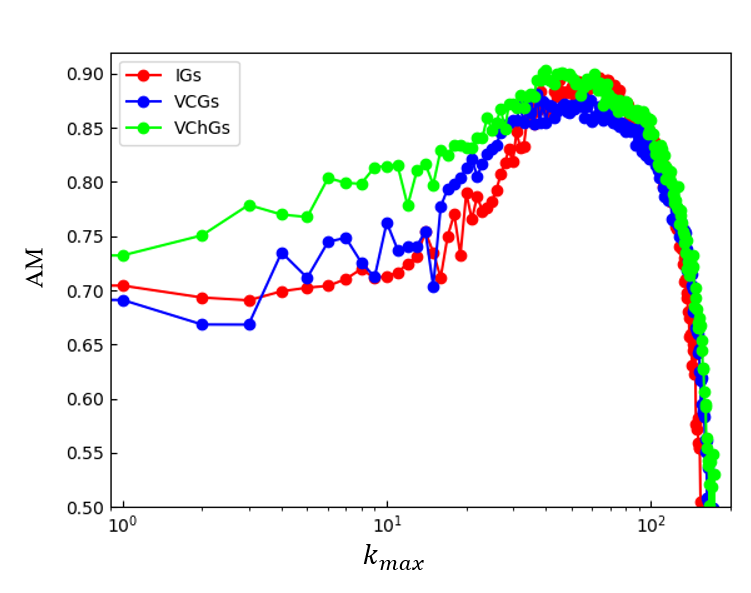}
    \caption{\label{fig:galfa_am}The correlation of AM and $k_{max}$. $k_{max}$ (the $x$-axis) indicates the spatial frequency from $k=0$ to $k=k_{max}$ is removed. The AM is calculated from IGs (red)/VCGs (blue)/VChGs (green) and the magnetic field is inferred from Planck polarization. The gradients are calculated from the GALFA-\ion{H}{1} data.}
\end{figure}

As shown in Fig.~\ref{fig:galfa}, when we use the \ion{H}{1} data within the full spatial frequency range, VChGs give good alignment with the Planck polarization is the low-intensity region (approximately $\rm I\le1200$ K m/s). As for the high-intensity region, there are more negative alignments, i.e., the rotated gradients are nearly perpendicular to the Planck polarization (see also \citealt{2020MNRAS.496.2868L}). One possible reason for the anti-alignment is the shock effect \citep{YL17b, IGs,H2}. Globally, we have AM = 0.73 for the VChGs and the Planck polarization. Similar to the numerical analysis above, we then remove the low-spatial frequency keeping only $k>50$ components in the velocity channel map. After removing low-spatial frequencies, we find the global AM between VChGs and the Planck polarization increases to 0.90. The majority of anti-alignments seen in high-intensity regions are suppressed. Also, in Fig.~\ref{fig:galfa_am}, we plot the correlation of AM and $k_{max}$ which indicates the spatial frequency from $k=0$ to $k=k_{max}$ is removed. For IGs, VCGs, and VChGs, we see the AM keeps a flat curve around value 0.70 until $k_{max}\approx15$. When $k_{max}\ge15$, the AM starts increasing to it maximum value $\approx$ 0.90 at $k_{max}\approx50$. The AM gets drops significantly if $k_{max}>100$. The results agree with our numerical studies above. To sum up, the most significant contribution to gradients comes from small-scale structures. The accuracy of GT can be improved by filtering out low-spatial frequencies.

\section{Discussion}
\label{sec:diss}

\subsection{Applying GT to interferometric data}
\label{sec:inter}
The measurement of magnetic fields is a hot topic in modern astrophysics. Owing to advances in dust grain alignment theory and synchrotron radiation theory, a comprehensive picture of POS magnetic field orientations has become more accessible using either dust polarization from background starlight or linearly polarized thermal dust emission or synchrotron emission. A number of these observations have been achieved by the ground telescopes/arrays \citep{1988AJ.....95..510B}, the balloon-borne telescopes (BLASTPol: \citealt{2014SPIE.9145E..0RG}), the airborne-platforms telescopes (SOFIA: \citealt{2019ApJ...882..113S}), and the space-satellite telescope (Planck: \citealt{2018arXiv180706212P}), spanning from millimeter, submillimeter to far-infrared (FIR) wavebands.

However, the measured spatial frequencies from interferometers are constrained by the baseline. In many cases, the single-dish observations necessary for obtaining low-spatial frequencies are missing. For instance, in \citet{2011Natur.478..214G}, the radio observations observed with the Australia Telescope Compact Array (ATCA) at a frequency of 1.4 GHz did not include single-dish measurements. \citet{2015A&A...583A.137J} observed the 3C 196 field with the Westerbork Synthesis Radio Telescope (WSRT) at 350 MHz. However, for the WSRT 350 MHz observations, a single-dish measurement at the same frequencies is not possible. Also, the radio data from LOw-Frequency ARray (LOFAR) suffers the loss of low-spatial-frequencies \citep{2014A&A...568A.101J}.

Our study shows that the missing low-spatial-frequencies in interferometers, however, benefit GT in terms of probing the magnetic fields. The GT is established on the basis of solid theory, i.e., anisotropic turbulence theory GS95 and turbulent LV99 reconnection. The theoretical prediction based on this theory is that magnetic field tracing is the best when only the spatial frequencies starting with $k=1/L_{block}$ are used (see \citealt{2020arXiv200207996L}). The other frequencies, in fact, may negatively interfere with the analysis. This special property of GT was earlier shown in \citet{2017ApJ...842...30L} for the SIGs and \citet{LY18a} for VChGs. It presents a remarkable ability to restore the entire map of magnetic fields only using the high spatial frequencies obtained via interferometric measurements. It should be noted that numerical simulations have limited inertial range, so our study does not include the case that the wavenumber is much larger than 100. However, in observation, the inertial range and turbulence's role are much extended to a large wavenumber range, so GT founded on the anisotropy of MHD turbulence works in large wavenumbers.

In this work, we show that the effect of missing low-spatial frequencies indeed improves the accuracy of intensity gradients (IGs, see \citealt{IGs}), velocity centroid gradients (VCGs, see \citealt{2017ApJ...835...41G}), velocity channel gradients (VChGs, see \citealt{LY18a}), synchrotron intensity gradients (SIGs, see \citealt{2017ApJ...842...30L}), and synchrotron polarization gradients (SPGs, see \citealt{2018ApJ...865...59L}). In particular, for subsonic turbulence, the (rotated) gradients give the best alignment with the magnetic fields in the range of spatial frequency $1/L_{block}<k<k_{diss}$, where $k_{diss}$ is the turbulence dissipation scale. 

As for super-sonic turbulence, the presence of shocks degrades the accuracy of intensity gradients \citep{YL17b,IGs}. As the magnetic fields in ISM tend to be perpendicular to the shock front \citep{2019ApJ...878..157X}, the jump condition of intensity will induce the gradients to be parallel to the magnetic fields rather than perpendicular. However, the shock effect can be suppressed by filtering out the low-spatial frequencies until the scale of high-density contrasts is less than the size of the sub-block. In addition, the gradients also flip their direction by 90$^\circ$ when the self-gravity is dominating over turbulence. The gradients are parallel to the magnetic field in the case of gravitational collapsing. Since the self-gravity spans both high and low spatial frequencies, it has a negative effect on the gradients on all scales. Nevertheless, this issue can be solved by first identifying the self-gravitating region and then re-rotating the gradient by 90$^\circ$ again \citep{survey,Hu20}.

\subsection{Obating correct interferometric data}
In addition to the missing low-spatial frequencies in an interferometer, several other aspects of interferometric observation need to be carefully considered. 
\subsubsection{Antenna primary beam attenuation}
Radio telescopes are inherently limited by their field of view, which is further affected by the attenuation of the antenna's primary beam. This attenuation can lead to a decrease in sensitivity at the edges of the field of view, potentially impacting the accuracy of GT. Mosaicking techniques \citep{1996A&AS..120..375S,2014MNRAS.444.2212B}, which involve combining multiple overlapping observations, can mitigate this limitation by extending the effective field of view and providing more uniform sensitivity.

\subsubsection{Limited Visibility Coverage}
Interferometers have restricted coverage in the visibility plane. This incomplete coverage results in the creation of "dirty images," which contain artifacts from the Fourier transform of the visibilities. These artifacts are characterized by the dirty beam, which describes the normalized response of the interferometer to a point source at the center of the image. To obtain an accurate representation of the sky, image reconstruction algorithms, such as the clean algorithm, must be employed \citep{2022PASP..134k4501C}. These algorithms iteratively remove the artifacts and recover the true image. 

\subsubsection{Background continuum sources}
Background point sources are not expected to correlate with the magnetic field in a turbulence medium. When using GT to trace magnetic fields, it is essential to remove the background point sources. 

Additionally, the presence of very bright continuum extragalactic radio sources can compromise the quality of observations near these sources due to their manifestation as extended emissions. This issue is particularly pronounced at very low frequencies, where such sources are abundant. While the SKA may mitigate this issue to some extent, the handling of interferometric data may need caution.

\subsection{Recovering 3D magnetic field structure}
A full 3D magnetic field model is crucial in many astrophysical problems, for instance, the removal of the galaxy foreground, star formation, and cosmic ray transportation. Such a 3D magnetic field model can be constructed by the GT.

For instance, the 3D galactic magnetic fields can be achieved through the galactic rotation curve and the measured atomic hydrogen distribution. \citet{2019ApJ...874...25G} produce the first 3D map of the galactic disk plane-of-the-sky magnetic field distribution using the velocity gradients. The authors successfully tested the resulting 3D model by comparing it with the measured stellar polarization. Also, \citet{velac} demonstrated the possibility of studying the 3D magnetic field POS distribution using multiple molecular tracers and their corresponding gradients (see also \citealt{PDF}). For example, by stacking the gradient maps calculated from $\rm ^{12}CO$, $\rm ^{13}CO$, and $\rm C^{18}O$ data, one could get the magnetic fields over three different optical depths providing spectroscopic information on gas dynamics up to a certain LOS depth. The Gradient Technique, therefore, give new insight into the readily available large-scale molecular line surveys, such as CHAMP \citep{2006JGRA..111.2304S} and ThRuMMs \citep{2015ApJ...812....7N}, or the neutral hydrogen atom distribution survey \citep{2016A&A...594A.116H}, and the COMPLETE survey \citep{2006AJ....131.2921R}.

Other important measurements are polarized synchrotron radiation and Faraday rotation \citep{2011MNRAS.412.2396F,2013pss5.book..641B,2015A&A...575A.118O,2016ApJ...830...38L,2017A&A...597A..98V}, which can reveal the magnetic fields of the Milky Way and neighboring galaxies. In addition, \citet{2018ApJ...865...59L} proposed to trace the 3D magnetic fields using SPGs. Due to the effect of Faraday depolarization, the measured synchrotron polarization can only come from the regions close to the observer. By varying the measurement wavelength, one can obtain the LOS depth of the synchrotron polarization (see Eq.~\ref{eq.11}). Similar to the multiple-tracers tomography discussed above, by stacking the SPGs maps corresponding to different LOS depths, a 3D magnetic field model can still be established \citep{2018ApJ...865...59L,2019ApJ...887..258H}. 
Also, SPGs show the possibility to trace the 3D magnetic fields in nearby galaxies. The Milky Way foreground distorts the direction of synchrotron polarization through the Faraday rotation. However, it does not contribute to the value of total polarization arising from an external galaxy. As a consequence, SPGs are not affected by the Faraday rotation foreground. The new instrument LOFAR \citep{2014A&A...568A.101J} provides an abundant low-frequency synchrotron data set, for which Faraday depolarization may be important. Also, the Square Kilometer Array (SKA) will provide detailed maps of diffuse emission with unprecedented resolution \citep{2015aska.confE..92J}. These new measurements of synchrotron emission definitely pave the way to study the magnetic fields using the GT. 

\subsection{Comparison with other studies}
As an output of modern MHD turbulence theories, the GT has been developed as a new method to trace magnetic fields. The first attempt to trace the magnetic fields using velocity centroid gradients (VCGs) was carried out by \citet{2017ApJ...835...41G}. The subsequent studies enable the usage of SIGs \citep{2017ApJ...842...30L}, IGs \citep{YL17b,IGs}, SPGs \citep{2018ApJ...865...59L}, and VChGs  \citep{LY18a}. This multiple-gradients ecosystem paves the way to study the magnetic fields orientation in transparent atomic gas \citep{YL17a,2019ApJ...874...25G, EB,SFA,2020MNRAS.496.2868L}, dense molecular gas \citep{survey, velac, 2020arXiv200715344A,2023MNRAS.524.4431H,2024MNRAS.530.1066L}, circumstellar medium \citep{2019ApJ...880..148G}, nearby galaxies \citep{galaxy,2023ApJ...946....8T,2023MNRAS.519.1068L,2024arXiv240402651Z}, and intracluster medium \citep{cluster,2024NatCo..15.1006H}. In addition to tracing the magnetic field direction, the GT also can be used to estimate the magnetization level \citep{2018ApJ...865...46L}, measure the sonic Mach number \citep{2018arXiv180200024Y}, distinguish shocks \citep{YL17b,IGs}, and identify gravitational collapsing regions \citep{Hu20}. In this work, we show the accuracy of the GT can be improved by removing low-spatial-frequencies, which is advantageous in using interferometric data.

Based on the alignment of \ion{H}{1} filaments with the magnetic fields, \citet{2014ApJ...789...82C} proposed the Rolling Hough Transform (RHT) to study the magnetic field in the diffuse interstellar region.
However, the RHT does not distinguish the perpendicular structures and the parallel structures with respect to the magnetic field. It is therefore incapable of revealing the magnetic field direction in shock regions \citep{2016MNRAS.460.1934M}. Furthermore, the output from RHT depends on three input parameters: smoothing kernel diameter, window diameter, and intensity threshold \citep{2014ApJ...789...82C,2020arXiv200715344A}. The adjustment of these inputs is empirically optimal when the reference polarization measurement is available. On the contrary, the GT does not require any adjustable parameters and can handle the shock effect with VChGs \citep{IGs}.

In particular, the VChGs is theoretically founded on the effect of velocity caustics \citep{LY18a}. This effect produces fluctuations from the field of velocity within thin channel maps, When the channel is sufficiently thin (see Eq.~\ref{eq1}), the turbulent velocity fluctuations dominate over intensity fluctuations arising from density clumping \citet{LP00}. In this case, the filamentary structures in thin velocity channels follow the statistics of velocity rather than intensity. 
\citet{2019ApJ...874..171C}, however, relate the \ion{H}{1} filamentary structures to the two-phase nature of \ion{H}{1}, while the role of non-linear spectroscopic mapping (i.e., the velocity caustics effect) was totally ignored. This interpretation limits the alignment of filaments in the thin channel maps with the magnetic fields to only \ion{H}{1}. However, the alignment has been reported by several studies in single-phase molecular clouds \citep{survey,velac,2020MNRAS.496.4546H}. The corresponding reply is presented in \citet{2019arXiv190403173Y} and \citet{2023MNRAS.524.2994H}. The discussion of the \ion{H}{1} filaments is out of the scope of this work and we will answer it elsewhere. 

In terms of practical applications, both the RHT and the GT approaches were used for tracing magnetic fields and predicting CMB foreground polarization arising from dust. For the latter task, the comparison of the performances of the two techniques was performed in \cite{2020MNRAS.496.2868L}, \citet{2020RNAAS...4..105H}, and \cite{2023MNRAS.524.2994H}. There it was shown that the GT can reproduce Planck polarization better than RHT, while not using any adjustable parameters. Our present study shows that the performance of the GT can be significantly improved, consolidating the lead of the GT approach. We note that the GT is a technique founded on theory and this provides the basis for further improvements. 

\begin{figure*}[t]
    \centering
    \includegraphics[width=1.0\linewidth]{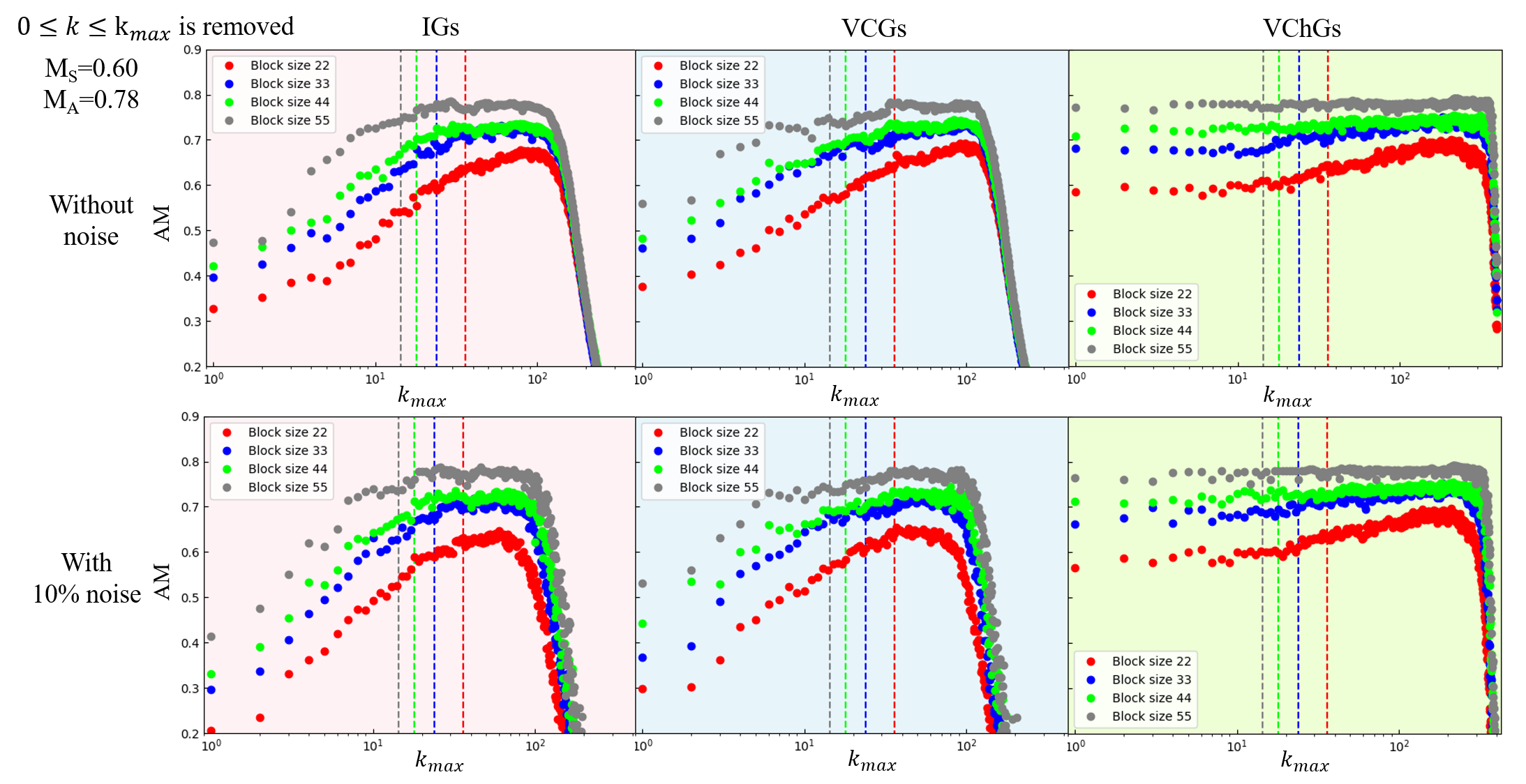}
    \caption{\label{fig:noise}The correlation of AM and $k_{max}$. $k_{max}$ (the x-axis) indicates the spatial frequency from $k=0$ to $k=k_{max}$ is removed. The AM is calculated from IGs (the 1st column)/VCGs (the 2nd column)/ VChGs (the 3rd column) and the magnetic field inferred from dust polarization. The dashed lines represent the corresponding $k$ values of the selected sub-block sizes. We used sub-sonic simulations $M_{\rm S}=0.60$ and $M_{\rm A}=0.78$ with the absence of noise (the 1st row) and the presence of noise (the 2nd row).}
\end{figure*}

\section{Summary}
\label{sec:conc}
 In this work, we studied the effect of spatial filtering on the accuracy of the GT mapping of the magnetic field. We have employed synthetic observations of sub-Alfv\'enic and trans-Alfv\'enic turbulence and GALFA HI survey data for our study. Our results can be briefly summarized in the following way: 
\begin{enumerate}
    \item For the sub-sonic environment, we show that filtering low spatial frequencies can improve the tracing of the magnetic field through the GT. The removal of high spatial frequencies gradually decreases the tracing.
    \item  Our application of the low K spatial filtering to a diffuse \ion{H}{1} from GALFA-\ion{H}{1} survey reveals a significant improvement in correspondence of GT predictions with the Planck 353 GHz polarization.
    \item We numerically demonstrate the possibility of using the GT with the interferometric data to study the galactic and extragalactic magnetic fields even when the full coverage of spatial frequencies is missing. 
 \end{enumerate}

\acknowledgments
Y.H. acknowledges the support for this work provided by NASA through the NASA Hubble Fellowship grant No. HST-HF2-51557.001 awarded by the Space Telescope Science Institute, which is operated by the Association of Universities for Research in Astronomy, Incorporated, under NASA contract NAS5-26555. A.L. acknowledges the support of the NSF grants AST 2307840 and ALMA SOSPADA-016. This publication utilizes data from Galactic ALFA \ion{H}{1} (GALFA -\ion{H}{1}) survey data set obtained with the Arecibo L-band Feed Array (ALFA) on the Arecibo 305m telescope. The Arecibo Observatory is operated by SRI International under a cooperative agreement with the National Science Foundation (AST-1100968), and in alliance with Ana G. Méndez-Universidad Metropolitana, and the Universities Space Research Association. The GALFA -\ion{H}{1} surveys have been funded by the NSF through grants to Columbia University, the University of Wisconsin, and the University of California. Based on observations obtained with Planck (http://www.esa.int/Planck), an ESA science mission with instruments and contributions directly funded by ESA Member States, NASA, and Canada.

%

\vspace{10mm}

\software{Julia \citep{2012arXiv1209.5145B}, ZEUS-MP/3D \citep{2006ApJS..165..188H}}

\appendix
\section{The effect of noise and resolution}
\label{appendx:noise}

\begin{figure*}[t]
    \subfigure[]{
    \centering
    \includegraphics[width=1.0\linewidth]{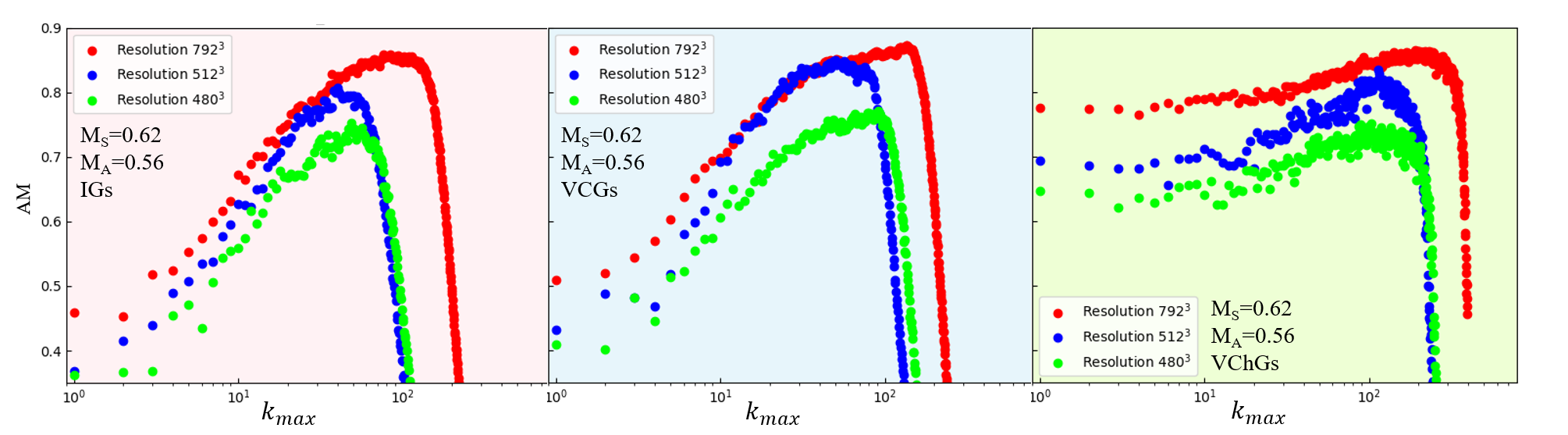}}
    \subfigure[]{
        \centering
    \includegraphics[width=1.0\linewidth]{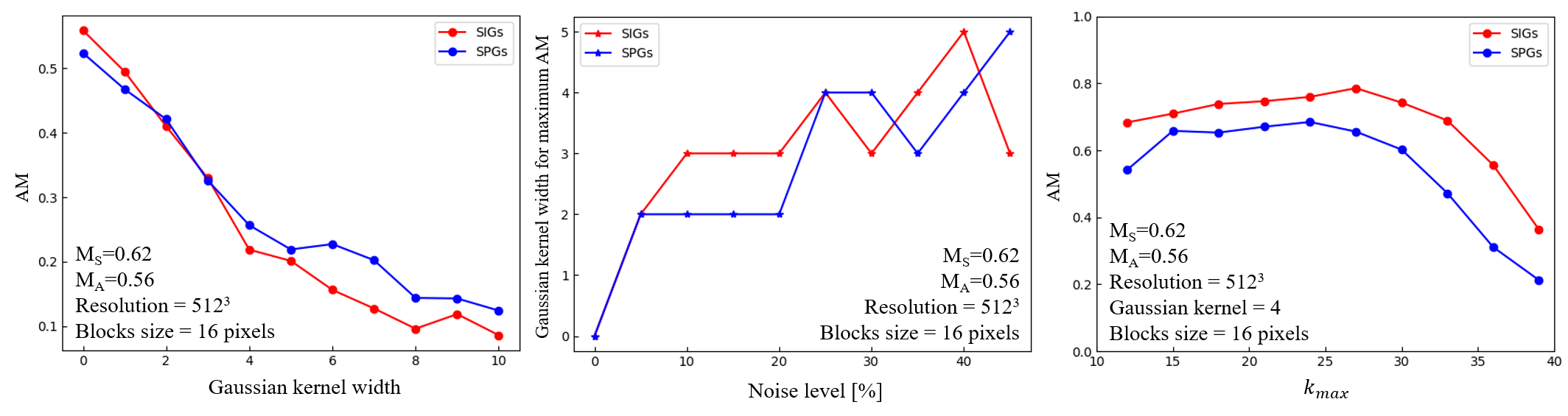}}
    \caption{\label{fig:gauss} \textbf{Panel a:} The correlation of AM and $k_{max}$. $k_{max}$ (the x-axis) indicates the spatial frequency from $k=0$ to $k=k_{max}$is removed. The AM is calculated from IGs (left)/VCGs (middle)/ VChGs (right) and the magnetic field is inferred from dust polarization.  \textbf{Panel b:} Left: plot of the AM as a function of Gaussian kernel width. The analysis is done for SIGs and SPGs in the full spatial range using simulation $M_{\rm S}=0.60$ and $M_{\rm A}=0.56$ in the absence of noise. Middle: the correlation between the Gaussian kernel width which outputs the maximum AM value and the noise level with respect to the mean intensity value. Right: the correlation of AM and $k_{max}$ which means the spatial frequencies from $k=0$ to $k_{max}$ is removed.}
\end{figure*}
\begin{figure*}[t]
    \centering
    \includegraphics[width=1.0\linewidth]{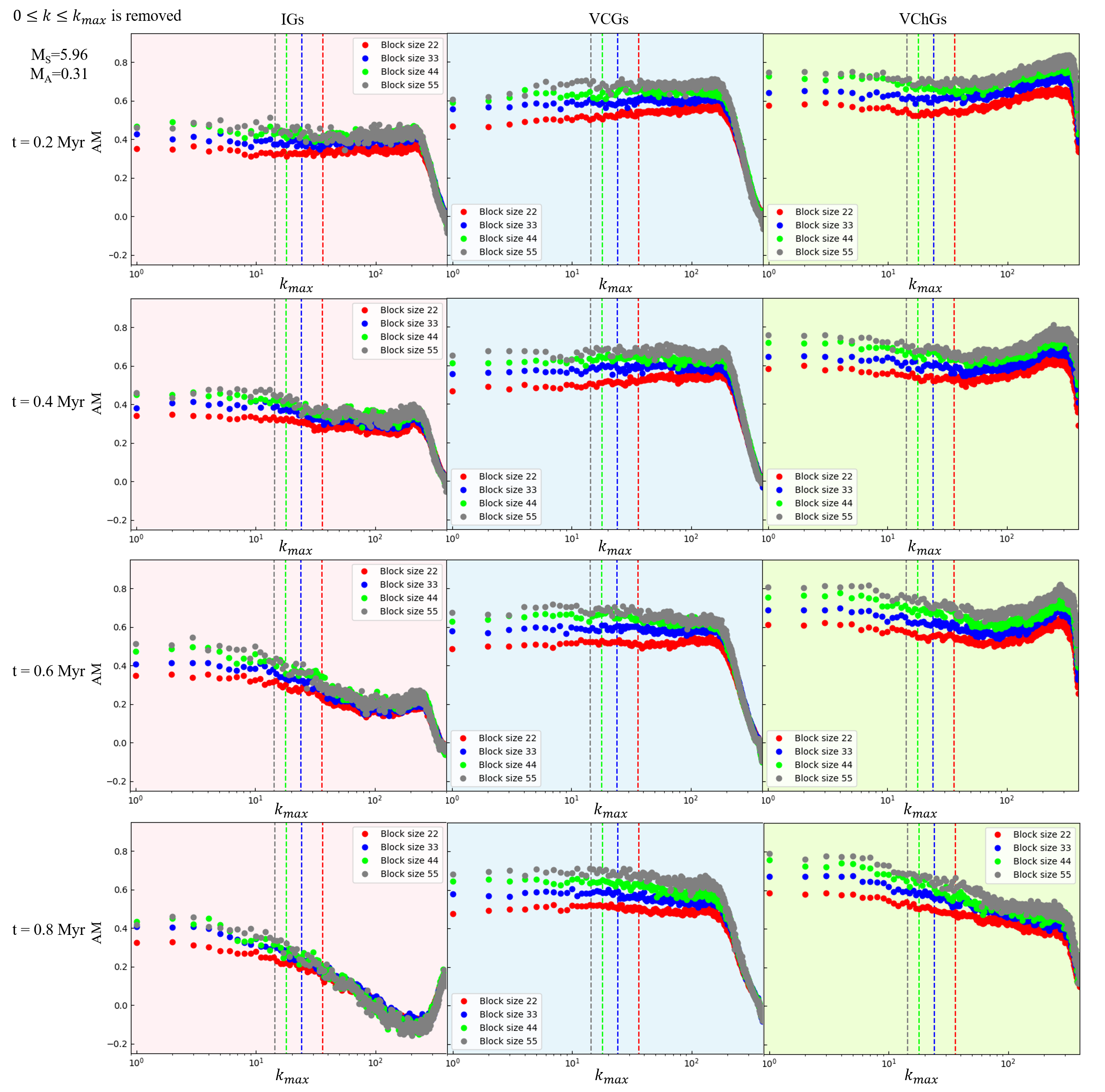}
    \caption{\label{fig:kmax_G} The correlation of AM and $k_{max}$. $k_{max}$ (the x-axis) indicates the spatial frequency from $k=0$ to $k=k_{max}$ is removed. The AM is calculated from IGs (the 1st column)/VCGs (the 2nd column)/ VChGs (the 3rd column) and the magnetic field inferred from dust polarization. The dashed lines represent the corresponding $k$ values of the selected sub-block sizes. We used supersonic simulations $M_{\rm S}=5.96$ and $M_{\rm A}=0.31$ with the presence of self-gravity here. $t$ represents the evolution time of the simulation after the self-gravity module is switched on.}
\end{figure*}
\begin{figure*}[t]
    \centering
    \includegraphics[width=1.0\linewidth]{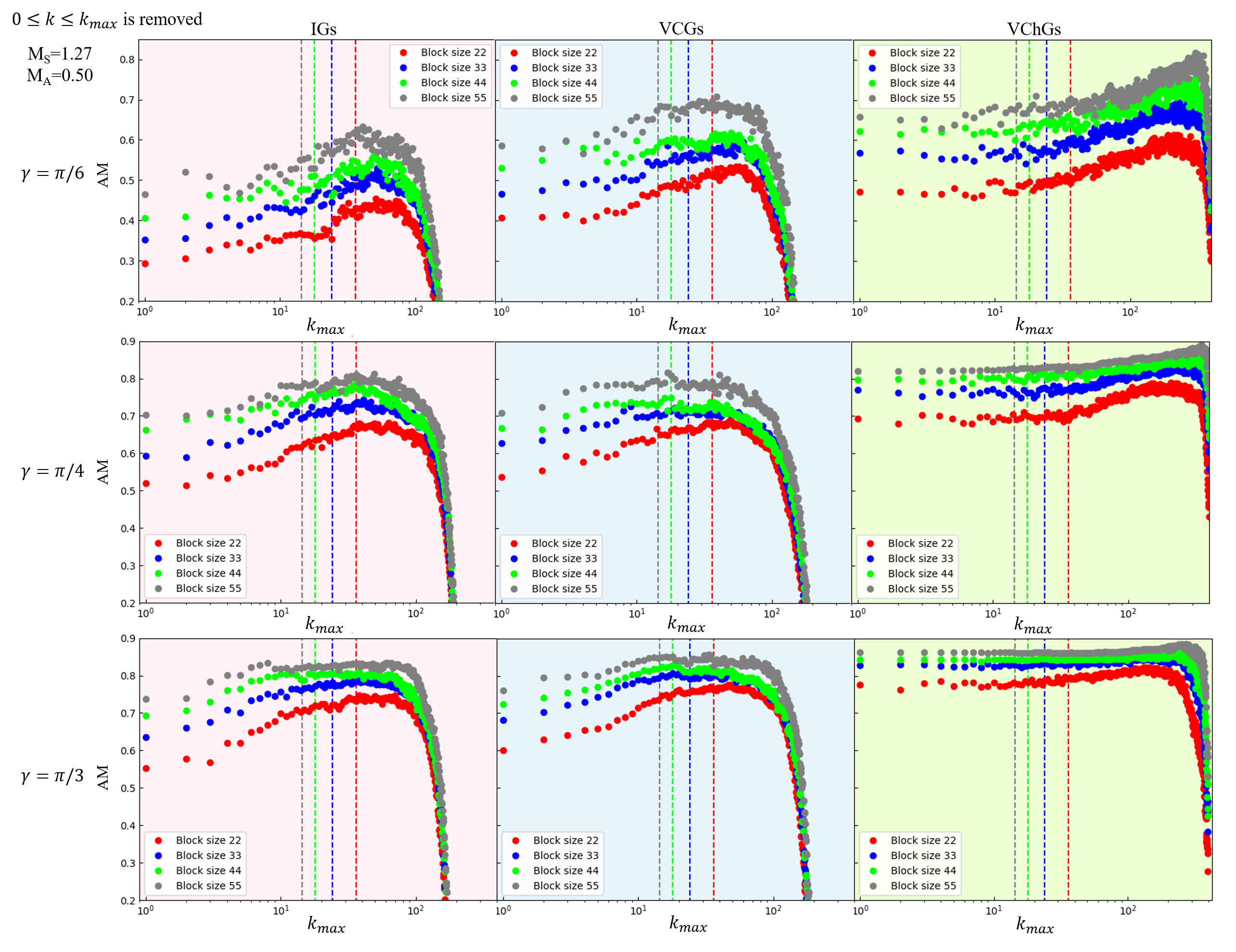}
    \caption{\label{fig:kmax_ang} The correlation of AM and $k_{max}$. $k_{max}$ (the x-axis) indicates the spatial frequency from $k=0$ to $k=k_{max}$ is removed. The AM is calculated from IGs (the 1st column)/VCGs (the 2nd column)/ VChGs (the 3rd column) and the magnetic field inferred from polarization. The dashed line represents the 3D dissipation scale $k\approx80$. We used supersonic simulations $M_{\rm S}=1.27$ and $M_{\rm A}=0.50$. $\gamma$ represents the inclination angle between the LOS and the magnetic fields.}
\end{figure*}
In Fig.~\ref{fig:noise}, we examine the performance of GT in the presence of noise. We produce 2D maps $I(x,y)$, $C(x,y)$, and $Ch(x,y)$ from sub-sonic simulations $M_{\rm S}=0.60$ and $M_{\rm A}=0.78$ and process them with the low-spatial frequency filter. Then we introduce Gaussian noise to the processed 2D maps. The noise's amplitude in each map is 10\% of their mean intensity. We find when the sub-block size is small ($\approx 22$ pixels) the noise degrades the maximum value of AM. However, the noise's contribution can be canceled out by increasing the sub-block size. Additionally, the VChGs is less sensitive to noise than IGs and VCGs.

As the dissipation of turbulence comes the numerical effect here, a simulation with high resolution usually guarantees a large dissipation scale $k_{diss}$. The scale $k_{diss}$ is crucial for the implementation of GT. Here then we analyze the effect coming from the finite resolution in numerical simulation. We vary the resolution of simulation A3 ($M_{\rm S}=0.60$ and $M_{\rm A}=0.56$) from 792$^3$, 512$^3$, and 480$^3$ and repeat the removing the low spatial frequency from $k=0$ to $k_{max}$ for IGs, VCGs, and VChGS. The results are presented in Fig.~\ref{fig:gauss}. We see that the removal of low spatial frequencies can also improve the alignment between gradients and the magnetic field weighted by density. The crucial numerical effect appears when $k_{max}$ is larger than the drop-off scale above which the AM starts decreasing. As discussed in \S~\ref{sec:result}, the scale is approximately the 2D dissipation scale of turbulence. Also, the maximum value of AM is slightly lower when the resolution is poor. 

The effect of missing spatial frequencies in SIGs was first studied by \citet{2017ApJ...842...30L}. There is a gradual decrease in the AM when low-spatial frequencies are removed, as a consequence of the Gaussian filtering applied to the numerical data. Indeed, the operation of a Gaussian filter with kernel width 4 is equivalent to convoluted images with a $(4^2+1)\times(4^2+1)$ matrix. The structures occupying less than 17 pixels are smoothed as a result. In Fig.~\ref{fig:gauss}, we analyze the correlation of SIGs' AM and the Gaussian kernel width using simulation $M_{\rm S}=0.60$ and $M_{\rm A}=0.56$, in the absence of noise. The AM here is calculated in terms of the magnetic field weighted by density. We find the AM is negatively proportional to the kernel width. It agrees with our consideration that only small-scale structures are crucial for the alignment but the Gaussian filter removes these structures. 

Additionally, we introduce noise to the synchrotron intensity map. Noise level $10\%$ means the Gaussian noise's magnitude is 10\% of the mean intensity value. We repeat the analysis of AM and the kernel width for different noise levels and find out the optimal kernel width outputting maximum AM. The result is shown in Fig.~\ref{fig:gauss}. We find that when the noise is about $10\%-30\%$, Gaussian kernel width 3 is a good choice to improve accuracy. Once the noise is more intensive than $30\%$, a larger kernel width is required. Also, we apply a Gaussian filter with kernel width 4 to the synchrotron intensity map and then plot the correlation between AM and $k_{max}$. $k_{max}$ means the spatial frequencies from $k=0$ to $k_{max}$ (i.e., low frequencies) is removed. Theoretically, structures smaller than 17 pixels will be removed by the filter, which corresponds to $k\approx30$ in our simulations. In Fig.~\ref{fig:gauss}, we observe that AM starts decreasing from $k_{max}\simeq45$, which agrees with the results shown in \citet{2017ApJ...842...30L}.

\section{The effect of self-gravity}
\label{appendx:gravity}
Initially, we consider non-self-gravitating conditions for all simulations. After the simulation is running for two sound crossing times, we switch on the self-gravity for simulation A7. We employ a periodic Fast Fourier Transform Poisson solver for the self-gravitating module and keep driving both turbulence and self-gravity until the evolution time $t\simeq 0.43$ $t_{ff}$, where $t_{ff} \simeq 1.88$ $\rm Myr$ is the free-fall time. The total mass $\rm M_{tot}$ in the simulated cubes A7 is $\rm M_{tot}\simeq8430.785 M_\odot,$ the magnetic Jean mass is $\rm M_{JB}\simeq86.95 M_\odot$, average magnetic field strength is $\rm B\simeq30.68 \mu G$, mass-to-flux ratio is $\Phi\simeq1.11$, and volume density is $\rm \rho\simeq315 cm^{-3}$ \citep{Hu20}. We repeat the analysis of removing low spatial frequency for the simulation at t = 0.2 Myr, 0.4 Myr, 0.6 Myr, and 0.8 Myr. We analyze only IGs, VCGs, and VChGs here. Once self-gravity takes place and the volume density increases, one would expect that most of the gas becomes molecular. In this case, the synchrotron emission would be not a good tracer for the GT.

Similar to \S.~\ref{sec:result}, in Fig.~\ref{fig:kmax_G}, we plot the AM value as a function of $k_{max}$, which means the spatial frequency from $k=0$ to $k=k_{max}$ is removed. We vary the sub-block size from 22 pixels to 55 pixels. At the beginning stage t = 0.2 Myr of self-gravitating, the AM values of IGs, VCGs, and VChGs perform similarly to the non-self-gravitating case (see Fig.~\ref{fig:sup_kmax}) as the turbulence is still dominating there. At t = 0.4 Myr, the AM of IGs starts decreasing in the range of $10\le k_{max}\le100$. VChGs also has similar behaviors. At later stages, t = 0.6 Myr and 0.8 Myr of self-gravitating, the decreasing trend of AM is more dramatic concerning $k_{max}$. The AM of VCGs also starts decreasing when $10\le k_{max}$. This anti-correlation of AM and $k_{max}$ comes from the self-gravity effect. \citet{Hu20} showed that in the case of gravitational collapse, as the acceleration induced by the collapse is along the magnetic field direction, density and velocity gradients are expected to change their orientation from perpendicular to magnetic fields to align with magnetic fields. In this case, the rotated gradients become perpendicular to the magnetic field resulting in a negative AM value. The effect of self-gravity is scale-dependent, which means the gravitational collapse can happen at a small local region where the gravitational energy surpasses the kinetic and magnetic field energy. Therefore, high-spatial frequency components give a lower or even negative AM value.

\section{Inclination angle between the LOS and the magnetic fields}
\label{appendx:gamma}
In the simulations, the mean magnetic field is initially perpendicular to the LOS. However, when the magnetic fields are inclined with respect to the LOS, we have to consider the projection effect for the projected magnetic field structures, which can get more irregular in the POS. Here we examine the effect introduced by the inclination angle to the k-space filter (see \S~\ref{sec:method}). We rotate the simulation cube so that the mean magnetic field has an inclination angle $\gamma=\pi/3$, $\pi/4$, and $\pi/6$ with respect to the LOS.

In Fig.~\ref{fig:kmax_ang}, we remove the spatial frequencies from $k=0$ to $k=k_{max}$ for the simulation $M_{\rm S}=1.27$ and $M_{\rm A}=0.50$. For IGs and VCGs, the AM values are still improved until $k_{max}$ achieves approximately the 3D dissipation scale $k\approx80$. When $k_{max}>80$, the AM significantly drops as the scale smaller than the dissipation scale contributes little to the alignment. The AM of VChGs is also improved when $k_{max}\le200$ and then starts significantly decreasing. As explained in \S~\ref{sec:result}, a larger drop-off scale of VChGs comes from the fact that the thin velocity channel has a flattened spectrum and a larger 2D dissipation scale. Nevertheless, we can see the alignment can be improved even if the inclination angle is less than $\pi/2$.
\begin{figure*}[t]
    \centering
    \includegraphics[width=1.0\linewidth]{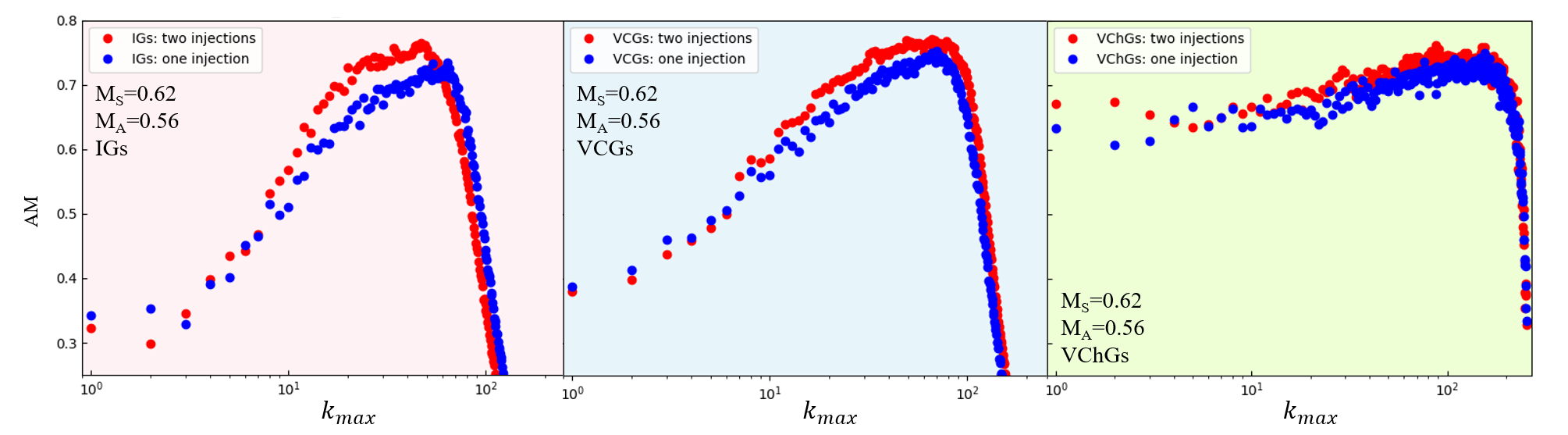}
    \caption{\label{fig:kmax_inject}The correlation of AM and $k_{max}$. $k_{max}$ (the x-axis) indicates the spatial frequency from $k=0$ to $k=k_{max}$ is removed. The AM is calculated from IGs (left)/VCGs (middle)/ VChGs (right) and the magnetic field weighted by density. The legend "two injections" indicates two numerical cubes are superimposed along the LOS, while "one injection" means only one numerical cube is used.}
\end{figure*}
\section{Multiple injections along the LOS}
\label{appendx:injection}
Here we examine the performance of the gradients when there are multiple injections along the LOS. Initially, the simulations are set so that there is only one injection scale on $k=2$. We then take two snapshots of the simulation cube in the same physical conditions ($M_{\rm S}=0.62$, $M_{\rm A}=0.56$, and resolution = 512$^3$) but at different evolution times. The two snapshots are superimposed along the LOS keeping the mean magnetic field still parallel to the POS. 

We repeat the analysis of $k_{max}$ as above, i.e., calculating the AM of IGs, VCGs, and VChGs after the spatial frequency from $k=0$ to $k=k_{max}$ is removed. We make a comparison of a single turbulence injection and two turbulence injections along the LOS. The results are presented in Fig.~\ref{fig:kmax_inject}. We can see the resulting AM exhibits a similar trend for the two cases as expected. Since the injection scales of the two snapshots are both at $k=2$, the resulting gradients are expected to be added up in a random walk manner. The case of multiple injection scales along the LOS will be studied elsewhere.
\begin{figure*}[p]
    \centering
    \includegraphics[width=1.0\linewidth,height=1.23\linewidth]{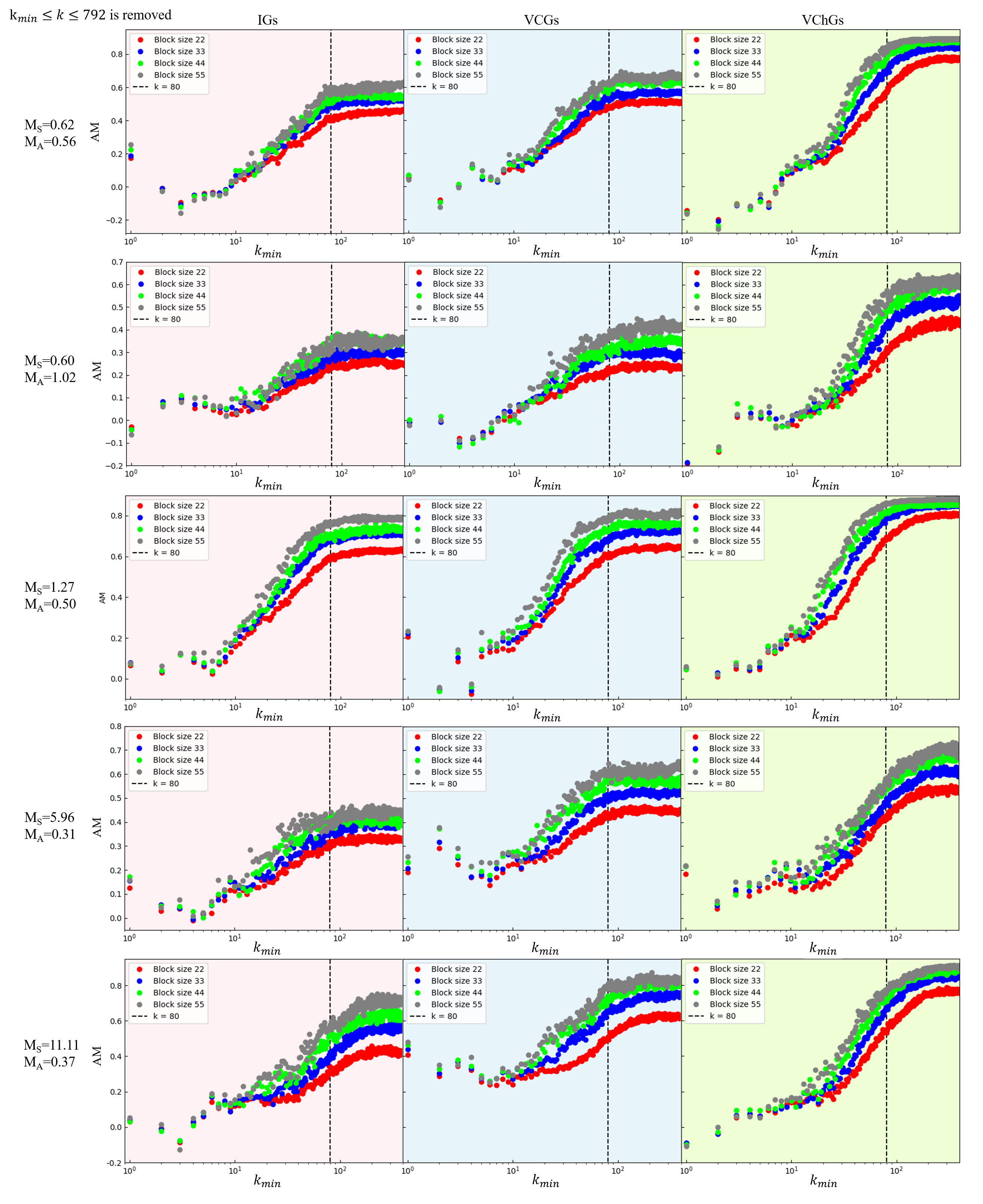}
    \caption{\label{fig:kmin} The correlation of AM and $k_{min}$. $k_{min}$ (the x-axis) indicates the spatial frequency from $k=k_{min}$ to $k=792$ is removed. The AM is calculated from IGs (the 1st column)/VCGs (the 2nd column)/VChGs (the 3rd column) and the magnetic field inferred from polarization. The dashed lines represent corresponding the dissipation scale $k_{diss}=80$.}
\end{figure*}
\section{Removing high-spatial frequency}
In this section, instead of removing low-spatial frequency, we test the effect of removing high-spatial frequency. In Fig.~\ref{fig:kmin}, we plot the AM value as a function of $k_{min}$, which means the spatial frequency from $k=k_{min}$ to $k=792$ which corresponds to 1 pixel is removed. 

Regarding both sub-sonic and supersonic turbulence, When $k_{min}$ is less than around 20, the AM values of IGs, VCGs, and VChGs are smaller than 0.2 with some variations. It means the low spatial frequency components, i.e., large-scale structures, have less contribution to the alignment of gradients and the magnetic fields. In the range of $20\le k_{min}\le80$, the AM values are monotonically increasing until the dissipation scale. When $100\le k_{min}$ which means the scales smaller than the dissipation scale ($\approx 80$) are removed, the AM values get saturated. 

To sum up, spatial frequency in the range of $0\le k \le15$ has less contribution to the alignment of gradients and magnetic fields. The spatial frequency in the range of $20\le k \le k_{diss}$, where $k_{diss}$ is the dissipation scale, is most crucial for tracing the magnetic field through gradients. 


\newpage
\begin{thebibliography}{}
\bibitem[Alina et al.(2020)]{2020arXiv200715344A} Alina, D., Montillaud, J., Hu, Y., et al.\ 2020, arXiv:2007.15344
\bibitem[Andersson et al.(2015)]{Aetal15} Andersson, B.-G., Lazarian, A., \& Vaillancourt, J.~E.\ 2015, \araa, 53, 501
\bibitem[Armstrong et al.(1995)]{1995ApJ...443..209A} Armstrong, J.~W., Rickett, B.~J., \& Spangler, S.~R.\ 1995, \apj, 443, 209
\bibitem[Ayachit et al.(2005)]{ayachit2015paraview} Ahrens, J., Geveci, B., Law, C.\ 205, Energy, 836, 717-732.
\bibitem[Balasubramanyam(2014)]{2014MNRAS.444.2212B} Balasubramanyam, R.\ 2014, \mnras, 444, 2212. 
\bibitem[Ballesteros-Paredes et al.(2007)]{2007prpl.conf...63B} Ballesteros-Paredes, J., Klessen, R.~S., Mac Low, M.-M., \& Vazquez-Semadeni, E.\ 2007, Protostars and Planets V, 63 
\bibitem[Barvainis et al.(1988)]{1988AJ.....95..510B} Barvainis, R., Clemens, D.~P., \& Leach, R.\ 1988, \aj, 95, 510
\bibitem[Beck \& Wielebinski(2013)]{2013pss5.book..641B} Beck, R. \& Wielebinski, R.\ 2013, Planets, Stars and Stellar Systems. Volume 5: Galactic Structure and Stellar Populations, 641
\bibitem[Beresnyak \& Lazarian (2019)]{BL19} Beresnyak, A., \& Lazarian, A.\ 2019, Turbulence in Magnetohydrodynamics, ~ISBN 3110262908.~De Gruyter Studies in Mathematical Physics (Book 12), April 1, 2019.
\bibitem[Beresnyak et al.(2005)]{2005ApJ...624L..93B} Beresnyak, A., Lazarian, A., \& Cho, J.\ 2005, \apjl, 624, L93
\bibitem[Bezanson et al.(2012)]{2012arXiv1209.5145B} Bezanson, J., Karpinski, S., Shah, V.~B., et al.\ 2012, arXiv e-prints, arXiv:1209.5145
\bibitem[Caprioli \& Spitkovsky(2014)]{2014ApJ...783...91C} Caprioli, D., \& Spitkovsky, A.\ 2014, \apj, 783, 91
\bibitem[Chepurnov \& Lazarian(2009)]{CL09} Chepurnov, A., \& Lazarian, A.\ 2009, \apj, 693, 1074 
\bibitem[Chepurnov \& Lazarian(2010)]{2010ApJ...710..853C} Chepurnov, A., \& Lazarian, A.\ 2010, \apj, 710, 853
\bibitem[Cho \& Lazarian(2002)]{CL02} Cho, J., \& Lazarian, A.\ 2002, Physical Review Letters, 88, 245001
\bibitem[Cho \& Lazarian(2003)]{2003MNRAS.345..325C} Cho, J., \& Lazarian, A.\ 2003, \mnras, 345, 325 
\bibitem[Cho \& Vishniac(2000)]{2000ApJ...539..273C} Cho, J., \& Vishniac, E.~T.\ 2000, \apj, 539, 273 
\bibitem[Cho et al.(2002)]{2002ApJ...564..291C} Cho, J., Lazarian, A., \& Vishniac, E.~T.\ 2002, \apj, 564, 291
\bibitem[Clark \& Hensley(2019)]{2019ApJ...887..136C} Clark, S.~E., \& Hensley, B.~S.\ 2019, \apj, 887, 136
\bibitem[Clark et al.(2014)]{2014ApJ...789...82C} Clark, S.~E., Peek, J.~E.~G., \& Putman, M.~E.\ 2014, \apj, 789, 82
\bibitem[Clark et al.(2019)]{2019ApJ...874..171C} Clark, S.~E., Peek, J.~E.~G., \& Miville-Desch{\^e}nes, M.-A.\ 2019, \apj, 874, 171
\bibitem[Clarke \& Ensslin(2006)]{2006AJ....131.2900C} Clarke, T.~E., \& Ensslin, T.~A.\ 2006, \aj, 131, 2900
\bibitem[Crutcher et al.(2010)]{2010ApJ...725..466C} Crutcher, R.~M., Wandelt, B., Heiles, C., et al.\ 2010, \apj, 725, 466
\bibitem[Crutcher(2012)]{2012ARAA...50..29} Crutcher, R. M.\ 2012, \ Annual Review of Astronomy and Astrophysics, 50, 29
\bibitem[CASA Team et al.(2022)]{2022PASP..134k4501C} CASA Team, Bean, B., Bhatnagar, S., et al.\ 2022, \pasp, 134, 114501. 
\bibitem[Dickey et al.(2019)]{2019ApJ...871..106D} Dickey, J.~M., Landecker, T.~L., Thomson, A.~J.~M., et al.\ 2019, \apj, 871, 106
\bibitem[Duan et al.(2021)]{2021ApJ...915L...8D} Duan, D., He, J., Bowen, T.~A., et al.\ 2021, \apjl, 915, L8.
\bibitem[Ekers \& Rots(1979)]{1979ASSL...76...61E} Ekers, R.~D. \& Rots, A.~H.\ 1979, IAU Colloq. 49: Image Formation from Coherence Functions in Astronomy, 61
\bibitem[Esquivel \& Lazarian(2005)]{2005ApJ...631..320E} Esquivel, A., \& Lazarian, A.\ 2005, \apj, 631, 320
\bibitem[Elmegreen \& Scalo(2004)]{2004ARA&A..42..211E} Elmegreen, B.~G., \& Scalo, J.\ 2004, \araa, 42, 211 
\bibitem[Fletcher et al.(2011)]{2011MNRAS.412.2396F} Fletcher, A., Beck, R., Shukurov, A., et al.\ 2011, \mnras, 412, 2396
\bibitem[Gaensler et al.(2011)]{2011Natur.478..214G} Gaensler, B.~M., Haverkorn, M., Burkhart, B., et al.\ 2011, \nat, 478, 214
\bibitem[Galitzki et al.(2014)]{2014SPIE.9145E..0RG} Galitzki, N., Ade, P.~A.~R., Angil{\`e}, F.~E., et al.\ 2014, \procspie, 9145, 91450R
\bibitem[Galli et al.(2006)]{2006ApJ...647..374G} Galli, D., Lizano, S., Shu, F.~H., \& Allen, A.\ 2006, \apj, 647, 374 
\bibitem[Goldreich \& Kylafis(1981)]{1981ApJ...243L..75G} Goldreich, P. \& Kylafis, N.~D.\ 1981, \apjl, 243, L75
\bibitem[Goldreich \& Kylafis(1982)]{1982ApJ...253..606G} Goldreich, P. \& Kylafis, N.~D.\ 1982, \apj, 253, 606
\bibitem[Goldreich \& Sridhar(1995)]{GS95} Goldreich, P., \& Sridhar, S.\ 1995, \apj, 438, 763 (GS95)
\bibitem[Gonz{\'a}lez-Casanova \& Lazarian(2017)]{2017ApJ...835...41G} Gonz{\'a}lez-Casanova, D.~F., \& Lazarian, A.\ 2017, \apj, 835, 41 
\bibitem[Gonz{\'a}lez-Casanova \& Lazarian(2019)]{2019ApJ...874...25G} Gonz{\'a}lez-Casanova, D.~F., \& Lazarian, A.\ 2019, \apj, 874, 25
\bibitem[Gonz{\'a}lez-Casanova \& Lazarian(2019)]{2019ApJ...880..148G} Gonz{\'a}lez-Casanova, D.~F. \& Lazarian, A.\ 2019, \apj, 880, 148
\bibitem[Haverkorn et al.(2006)]{2006ApJS..167..230H} Haverkorn, M., Gaensler, B.~M., McClure-Griffiths, N.~M., et al.\ 2006, \apjs, 167, 230
\bibitem[Hayes et al.(2006)]{2006ApJS..165..188H} Hayes, J.~C., Norman, M.~L., Fiedler, R.~A., et al.\ 2006, \apjs, 165, 188
\bibitem[Heiles(1976)]{1976ARA&A..14....1H} Heiles, C.\ 1976, \araa, 14, 1
\bibitem[Heyer et al.(2020)]{2020MNRAS.496.4546H} Heyer, M., Soler, J.~D., \& Burkhart, B.\ 2020, \mnras, 496, 4546
\bibitem[HI4PI Collaboration et al.(2016)]{2016A&A...594A.116H} HI4PI Collaboration, Ben Bekhti, N., Flöer, L., et al.\ 2016, \aap, 594, A116
\bibitem[Higdon(1984)]{1984ApJ...285..109H} Higdon, J.~C.\ 1984, \apj, 285, 109
\bibitem[Hsieh et al.(2019)]{2019ApJ...873...16H} Hsieh, C.-H., Hu, Y., Lai, S.-P., et al.\ 2019, \apj, 873, 16.
\bibitem[Hu \& Lazarian(2020a)]{2020RNAAS...4..105H} Hu, Y., \& Lazarian, A.\ 2020, Research Notes of the American Astronomical Society, 4, 105
\bibitem[Hu \& Lazarian(2021)]{PDF} Hu, Y. \& Lazarian, A.\ 2021, \mnras, 502, 1768. 
\bibitem[Hu et al.(2018)]{PCA} Hu, Y., Yuen, K.~H., \& Lazarian, A.\ 2018, \mnras, 480, 1333.
\bibitem[Hu et al.(2019a)]{survey}  Hu, Y., Yuen, K.~H., Lazarian, V., et al.\ 2019, \href{https://doi.org/10.1038/s41550-019-0769-0}{Nature Astronomy, 3, 776.}
\bibitem[Hu et al.(2019b)]{velac} Hu, Y., Yuen, K.~H., Lazarian, A., et al.\ 2019, \apj, 884, 137
\bibitem[Hu et al.(2019c)]{IGs} Hu, Y., Yuen, K.~H., \& Lazarian, A.\ 2019, \apj, 886, 17
\bibitem[Hu et al.(2020a)]{EB} Hu, Y., Yuen, K.~H., \& Lazarian, A.\ 2020, \apj, 888, 96
\bibitem[Hu et al.(2020b)]{Hu20} Hu, Y., Lazarian, A., \& Yuen, K.~H.\ 2020, \apj, 897, 123
\bibitem[Hu et al.(2020c)]{cluster} Hu, Y., Lazarian, A., Li, Y., et al.\ 2020, \apj, 901, 162.
\bibitem[Hu et al.(2020d)]{H2} Hu, Y., Lazarian, A., \& Bialy, S.\ 2020, \apj, 905, 129
\bibitem[Hu et al.(2021a)]{SFA} Hu, Y., Xu, S., \& Lazarian, A.\ 2021, \apj, 911, 37. 
\bibitem[Hu et al.(2021b)]{2021ApJ...915...67H} Hu, Y., Lazarian, A., \& Xu, S.\ 2021, \apj, 915, 67. 
\bibitem[Hu et al.(2022a)]{CMZ1} Hu, Y., Lazarian, A., \& Wang, Q.~D.\ 2022, \mnras, 511, 829. 
\bibitem[Hu et al.(2022b)]{CMZ2} Hu, Y., Lazarian, A., \& Wang, Q.~D.\ 2022, \mnras, 513, 3493. 
\bibitem[Hu et al.(2022c)]{galaxy} Hu, Y., Lazarian, A., Beck, R., et al.\ 2022, arXiv:2206.05423
\bibitem[Hu et al.(2023)]{2023MNRAS.524.2994H} Hu, Y., Lazarian, A., Alina, D., et al.\ 2023, \mnras, 524, 2994. 
\bibitem[Hu \& Lazarian(2023)]{2023MNRAS.524.4431H} Hu, Y. \& Lazarian, A.\ 2023, \mnras, 524, 4431. 
\bibitem[Hu et al.(2024)]{2024NatCo..15.1006H} Hu, Y., Stuardi, C., Lazarian, A., et al.\ 2024, Nature Communications, 15, 1006. 
\bibitem[Ho et al.(2019)]{2019ApJ...887..258H} Ho, K.~W., Yuen, K.~H., Leung, P.~K., et al.\ 2019, \apj, 887, 258
\bibitem[Jeli{\'c} et al.(2014)]{2014A&A...568A.101J} Jeli{\'c}, V., de Bruyn, A.~G., Mevius, M., et al.\ 2014, \aap, 568, A101
\bibitem[Jeli{\'c} et al.(2015)]{2015A&A...583A.137J} Jeli{\'c}, V., de Bruyn, A.~G., Pandey, V.~N., et al.\ 2015, \aap, 583, A137
\bibitem[Johnston-Hollitt et al.(2015)]{2015aska.confE..92J} Johnston-Hollitt, M., Govoni, F., Beck, R., et al.\ 2015, Advancing Astrophysics with the Square Kilometre Array (AASKA14), 92
\bibitem[Kandel et al.(2016)]{KLP16} Kandel, D., Lazarian, A., \& Pogosyan, D.\ 2016, \mnras, 461, 1227
\bibitem[Kowal et al.(2007)]{2007ApJ...658..423K} Kowal, G., Lazarian, A., \& Beresnyak, A.\ 2007, \apj, 658, 423 
\bibitem[Kronberg(1994)]{1994RPPh...57..325K} Kronberg, P.~P.\ 1994, Reports on Progress in Physics, 57, 325
\bibitem[Larson(1981)]{1981MNRAS.194..809L} Larson, R.~B.\ 1981, \mnras, 194, 809 
\bibitem[Lazarian(2006)]{2006ApJ...645L..25L} Lazarian, A.\ 2006, \apjl, 645, L25
\bibitem[Lazarian \& Hoang(2007)]{2007ApJ...669L..77L} Lazarian, A., \& Hoang, T.\ 2007, \apjl, 669, L77
\bibitem[Lazarian \& Pogosyan(2000)]{LP00} Lazarian, A., \& Pogosyan, D.\ 2000, \apj, 537, 720 (LP00)
\bibitem[Lazarian \& Pogosyan(2004)]{LP04} Lazarian, A. \& Pogosyan, D.\ 2004, \apj, 616, 943.
\bibitem[Lazarian \& Pogosyan(2008)]{2008ApJ...686..350L} Lazarian, A., \& Pogosyan, D.\ 2008, \apj, 686, 350
\bibitem[Lazarian \& Pogosyan(2012)]{2012ApJ...747....5L} Lazarian, A., \& Pogosyan, D.\ 2012, \apj, 747, 5
\bibitem[Lazarian \& Pogosyan(2016)]{LP16} Lazarian, A. \& Pogosyan, D.\ 2016, \apj, 818, 178
\bibitem[Lazarian \& Vishniac(1999)]{LV99} Lazarian, A., \& Vishniac, E.~T.\ 1999, \apj, 517, 700 
\bibitem[Lazarian \& Yuen(2018a)]{LY18a} Lazarian, A., \& Yuen, K.~H.\ 2018a, \apj, 853, 96 (LV99)
\bibitem[Lazarian \& Yuen(2018b)]{2018ApJ...865...59L} Lazarian, A., \& Yuen, K.~H.\ 2018b, \apj, 865, 59.
\bibitem[Lazarian et al.(2001)]{2001ApJ...555..130L} Lazarian, A., Pogosyan, D., V{\'a}zquez-Semadeni, E., et al.\ 2001, \apj, 555, 130
\bibitem[Lazarian et al.(2017)]{2017ApJ...842...30L} Lazarian, A., Yuen, K.~H., Lee, H., et al.\ 2017, \apj, 842, 30
\bibitem[Lazarian et al.(2018a)]{2018ApJ...865...46L} Lazarian, A., Yuen, K.~H., Ho, K.~W., et al.\ 2018, \apj, 865, 46.
\bibitem[Lazarian et al.(2020a)]{2020arXiv200207996L} Lazarian, A., Yuen, K.~H., \& Pogosyan, D.\ 2020, arXiv e-prints, arXiv:2002.07996
\bibitem[Lazarian(2007)]{2007JQSRT.106..225L} Lazarian, A.\ 2007, \jqsrt, 106, 225
\bibitem[Lenc et al.(2016)]{2016ApJ...830...38L} Lenc, E., Gaensler, B.~M., Sun, X.~H., et al.\ 2016, \apj, 830, 38
\bibitem[Li \& Henning(2011)]{2011Natur.479..499L} Li, H.-B., \& Henning, T.\ 2011, \nat, 479, 499 
\bibitem[Liu et al.(2022)]{2022MNRAS.510.4952L} Liu, M., Hu, Y., \& Lazarian, A.\ 2022, \mnras, 510, 4952. 
\bibitem[Liu et al.(2023)]{2023MNRAS.519.1068L} Liu, M., Hu, Y., Lazarian, A., et al.\ 2023, \mnras, 519, 1068. 
\bibitem[Liu et al.(2024)]{2024MNRAS.530.1066L} Liu, M., Hu, Y., \& Lazarian, A.\ 2024, \mnras, 530, 1066. 
\bibitem[Lu et al.(2020)]{2020MNRAS.496.2868L} Lu, Z., Lazarian, A., \& Pogosyan, D.\ 2020, \mnras, 496, 2868
\bibitem[Mac Low \& Klessen(2004)]{2004RvMP...76..125M} Mac Low, M.-M., \& Klessen, R.~S.\ 2004, Reviews of Modern Physics, 76, 125
\bibitem[Malinen et al.(2016)]{2016MNRAS.460.1934M} Malinen, J., Montier, L., Montillaud, J., et al.\ 2016, \mnras, 460, 1934
\bibitem[Matteini et al.(2020)]{2020FrASS...7...83M} Matteini, L., Franci, L., Alexandrova, O., et al.\ 2020, Frontiers in Astronomy and Space Sciences, 7, 83.
\bibitem[Maron \& Goldreich(2001)]{2001ApJ...554.1175M} Maron, J., \& Goldreich, P.\ 2001, \apj, 554, 1175
\bibitem[Matthaeus et al.(1983)]{1983PhRvL..51.1484M} Matthaeus, W.~H., Montgomery, D.~C., \& Goldstein, M.~L.\ 1983, \prl, 51, 1484
\bibitem[McKee \& Ostriker(2007)]{MO07} McKee, C.~F., \& Ostriker, E.~C.\ 2007, \araa, 45, 565 
\bibitem[Montgomery \& Turner(1981)]{1981PhFl...24..825M} Montgomery, D., \& Turner, L.\ 1981, Physics of Fluids, 24, 825
\bibitem[Nguyen et al.(2015)]{2015ApJ...812....7N} Nguyen, H., Nguyen Lu'o'ng, Q., Martin, P.~G., et al.\ 2015, \apj, 812, 7
\bibitem[Oppermann et al.(2012)]{2012A&A...542A..93O} Oppermann, N., Junklewitz, H., Robbers, G., et al.\ 2012, \aap, 542, A93
\bibitem[Oppermann et al.(2015)]{2015A&A...575A.118O} Oppermann, N., Junklewitz, H., Greiner, M., et al.\ 2015, \aap, 575, A118
\bibitem[Peek et al.(2018)]{2018ApJS..234....2P} Peek, J.~E.~G., Babler, B.~L., Zheng, Y., et al.\ 2018, \apjs, 234, 2
\bibitem[Pingel et al.(2018)]{2018ApJ...865...36P} Pingel, N.~M., Pisano, D.~J., Heald, G., et al.\ 2018, \apj, 865, 36
\bibitem[Planck Collaboration et al.(2016)]{2016A&A...596A.103P} Planck Collaboration, Adam, R., Ade, P.~A.~R., et al.\ 2016, \aap, 596, A103.
\bibitem[Planck Collaboration et al.(2020a)]{2018arXiv180706212P} Planck Collaboration, Aghanim, N., Akrami, Y., et al.\ 2020, \aap, 641, A12. 
\bibitem[Planck Collaboration et al.(2020b)]{2018arXiv180706207P} Planck Collaboration, Aghanim, N., Akrami, Y., et al.\ 2020, \aap, 641, A3. 
\bibitem[Rau \& Cornwell(2011)]{2011A&A...532A..71R} Rau, U. \& Cornwell, T.~J.\ 2011, \aap, 532, A71
\bibitem[Ridge et al.(2006)]{2006AJ....131.2921R} Ridge, N.~A., Di Francesco, J., Kirk, H., et al.\ 2006, \aj, 131, 2921
\bibitem[Santos et al.(2019)]{2019ApJ...882..113S} Santos, F.~P., Chuss, D.~T., Dowell, C.~D., et al.\ 2019, \apj, 882, 113
\bibitem[Sault et al.(1996)]{1996A&AS..120..375S} Sault, R.~J., Staveley-Smith, L., \& Brouw, W.~N.\ 1996, \aaps, 120, 375
\bibitem[Schmaltz et al.(2024)]{2024MNRAS.528.3897S} Schmaltz, T., Hu, Y., \& Lazarian, A.\ 2024, \mnras, 528, 3897. 
\bibitem[Shebalin et al.(1983)]{1983JPlPh..29..525S} Shebalin, J.~V., Matthaeus, W.~H., \& Montgomery, D.\ 1983, Journal of Plasma Physics, 29, 525
\bibitem[Simard-Normandin et al.(1981)]{1981ApJS...45...97S} Simard-Normandin, M., Kronberg, P.~P., \& Button, S.\ 1981, \apjs, 45, 97
\bibitem[Soler et al.(2013)]{Soler2013} Soler, J.~D., Hennebelle, P., Martin, P.~G., et al.\ 2013, \apj, 774, 128
\bibitem[Stolle et al.(2006)]{2006JGRA..111.2304S} Stolle, C., L{\"u}hr, H., Rother, M., et al.\ 2006, Journal of Geophysical Research (Space Physics), 111, A02304
\bibitem[Tram et al.(2023)]{2023ApJ...946....8T} Tram, L.~N., Bonne, L., Hu, Y., et al.\ 2023, \apj, 946, 8. 
\bibitem[Van Eck et al.(2017)]{2017A&A...597A..98V} Van Eck, C.~L., Haverkorn, M., Alves, M.~I.~R., et al.\ 2017, \aap, 597, A98
\bibitem[Wang et al.(2016)]{2016ApJ...816...15W} Wang, X., Tu, C., Marsch, E., et al.\ 2016, \apj, 816, 15. 
\bibitem[Widrow(2002)]{2002RvMP...74..775W} Widrow, L.~M.\ 2002, Reviews of Modern Physics, 74, 775
\bibitem[Xu et al.(2019)]{2019ApJ...878..157X} Xu, S., Ji, S., \& Lazarian, A.\ 2019, \apj, 878, 157
\bibitem[Xu \& Hu(2021)]{2021ApJ...910...88X} Xu, S. \& Hu, Y.\ 2021, \apj, 910, 88. 
\bibitem[Yan \& Lazarian(2006)]{2006ApJ...653.1292Y} Yan, H. \& Lazarian, A.\ 2006, \apj, 653, 1292
\bibitem[Yan \& Lazarian(2007)]{2007ApJ...657..618Y} Yan, H. \& Lazarian, A.\ 2007, \apj, 657, 618
\bibitem[Yan \& Lazarian(2008)]{2008ApJ...677.1401Y} Yan, H. \& Lazarian, A.\ 2008, \apj, 677, 1401
\bibitem[Yan \& Lazarian(2015)]{2015ASSL..407...89Y} Yan, H. \& Lazarian, A.\ 2015, Magnetic Fields in Diffuse Media, 89
\bibitem[Yuen \& Lazarian(2017a)]{YL17a} Yuen, K.~H., \& Lazarian, A.\ 2017, \apjl, 837, L24  
\bibitem[Yuen \& Lazarian(2017b)]{YL17b} Yuen, K.~H., \& Lazarian, A.\ 2017, arXiv:1703.03026
\bibitem[Yuen \& Lazarian(2020)]{2018arXiv180200024Y} Yuen, K.~H., \& Lazarian, A.\ 2018, arXiv e-prints, arXiv:1802.00024
\bibitem[Yuen et al.(2018)]{2018ApJ...865...54Y} Yuen, K.~H., Chen, J., Hu, Y., et al.\ 2018, \apj, 865, 54 
\bibitem[Yuen et al.(2019)]{2019arXiv190403173Y} Yuen, K.~H., Hu, Y., Lazarian, A., et al.\ 2019, arXiv e-prints , arXiv:1904.03173.
\bibitem[Yusef-Zadeh et al.(2024)]{2024MNRAS.527.1275Y} Yusef-Zadeh, F., Wardle, M., Arendt, R., et al.\ 2024, \mnras, 527, 1275.
\bibitem[Zhao et al.(2022)]{2022ApJ...934...45Z} Zhao, M., Zhou, J., Hu, Y., et al.\ 2022, \apj, 934, 45. 
\bibitem[Zhao et al.(2024)]{2024ApJ...962...89Z} Zhao, S., Yan, H., Liu, T.~Z., et al.\ 2024, \apj, 962, 89. 
\bibitem[Zhao et al.(2024)]{2024arXiv240402651Z} Zhao, M., Zhou, J., Baan, W.~A., et al.\ 2024, \apj in print, arXiv:2404.02651. 
\bibitem[Zhang et al.(2019)]{2019MNRAS.486.4813Z} Zhang, J.-F., Lazarian, A., Ho, K.~W., et al.\ 2019, \mnras, 486, 4813
\bibitem[Zhang \& Yan(2018)]{2018MNRAS.475.2415Z} Zhang, H. \& Yan, H.\ 2018, \mnras, 475, 2415
\end{thebibliography}



\end{document}